\documentstyle[aps,psfig,epsf]{revtex}

\begin{document}

\def\overlay#1#2{\setbox0=\hbox{#1}\setbox1=\hbox to \wd0{\hss #2\hss}#1
-2\wd0\copy1}

\twocolumn[\hsize\textwidth\columnwidth\hsize\csname@twocolumnfalse\endcsname
\title  {Critical currents in Josephson junctions  with macroscopic
defects}
\author{N. Stefanakis, N. Flytzanis}
\address{ Department of Physics, University of Crete,
 P.O. Box 2208, GR-71003, Heraklion, Crete, Greece}
\date{\today}
\maketitle
\begin{abstract}
\begin{center}
\parbox{14cm}
{

The critical currents  in Josephson junctions of
conventional superconductors with macroscopic defects are
calculated for different defect critical current densities as a
function of the magnetic field. We also study the evolution of the
different modes with the defect position, at zero external field. We
study the stability of the solutions and derive simple arguments,
that could help the defect characterization. In most cases a reentrant
behavior is seen, where both a maximum and a minimum current exist.
}
\end{center}
\end{abstract}
\vskip2pc]

\tighten

\section{Introduction} The interaction of localized magnetic flux
(fluxons) with defects (natural or artificial) or impurities in
superconductors or junctions has an important effect in the properties
of bulk superconductors or the behavior of Josephson junctions
correspondingly \cite{conf}. The flux trapping from defects which is
of major importance in Josephson junctions \cite{barone} can modify
the properties of polycrystalline materials with physical
dislocations, for example grain boundary junctions \cite{dimos}. In
this category one can also consider grain boundary junctions in
YBa$_2$Cu$_3$O$_7$ \cite{sarnelli} where the tunneling current is a
strongly varying function along the boundary. This strong
inhomogeneity makes them good candidates for SQUID type structures
\cite{gross1}. Phenomenologically the current-voltage characteristics
of grain boundary junctions are well described \cite{gross2} by the
resistively shunted junction model \cite{mccumber}.
The grain boundary lines often tend to curve, while the junction is
very inhomogeneous and contains nonsuperconducting impurities and
facets of different length scales\cite{miller,ayache}.
The linear increase of the critical current with length in grain boundary
junctions with high-$T_c$ superconductors, which is  a different
behavior from the saturation in the inline geometry of a perfect
junction, can be explained by the presence of impurities\cite{fehren}.
Therefore it is interesting to study flux trapping in impurities
especially when it can be controlled. Modern fabrication techniques
can with relative ease engineer any defect configuration in an
extremely controlled way.

In bulk materials are several types of defects that can
influence the critical current in high temperature superconductors
like YBa$_2$Cu$_3$O$_x$ materials. They include 3d
inclusions, 2d grain boundaries and twin boundaries, and point defects
like dopants substitutions, oxygen vacancies\cite{conf}. For example
the homogeneous precipitation of fine Y$_2$BaCuO$_5$ non-superconducting
particles in the melt processing of YBa$_2$Cu$_3$O$_x$ leads to high $J_c$
values due to the particle pinning centers\cite{oka}.
 Similar behavior is observed in NdBa$_2$Cu$_3$O$_x$ bulk crystals with
Nd$_4$Ba$_2$Cu$_2$O$_{10}$ particles\cite{takagi}. Of interest is also the
 case of the peak effect in twin-free $Y123$ with oxygen deficiency.
 In this case, one  sees a linear increase (peak effect) of the
 critical current at small magnetic fields, when growth is under
 oxygen reduction\cite{wolf}. For the fully oxidized crystal one
 expects a decrease. The peak effect is attributed to flux trapping.
 Information on the defect density and activation energies  can also
 be obtained from the I-V characteristics, as was the case for several
 types of defects which were also compared to $Au^+$ irradiated
 samples with artificial columnar defects\cite{camerlingo}. These
 columnar defects also act to trap flux lines in an YBCO film which is
 considered as a network of intergrain Josephson junctions modulated
 by the defects. In this case assuming a distribution of contact
 lengths one finds a plateau in the critical current density vs. the
 logarithm of the field\cite{mezzetti}.

The study of long size of impurities is going to give information
 beyond theories which concern small amplitude of inhomogeneities
 \cite{virocur}. Also it is possible for a direct comparison of the
 numerical results with experiments in long junctions obtained with
 electron beam lithography \cite{kroger}. This is a powerful technique
 which allows the preparation and control of arrays of pinning
 centers. Another method is the ionic irradiation which produces a
 particular kind of disordered arrays, consisting of nanosized
 columnar defects \cite{camerlingo,beek}.  The variation of the
 critical current density can also occur due to temperature gradients
 \cite{krasnov}.

   The activity in the area of high critical current
 densities in the presence of a magnetic field is hampered by defects
 due to the difficulty of having a high quality junction with a very
 thin intermediate layer. Thus significant activity has been devoted,
 since for example the energy resolution of the SQUID
\cite{ketchen} and the maximum operating frequency of the single
flux quantum logic circuit \cite{likharev}, to name a few
applications,  depend inversely and directly respectively on the plasma
frequency  $\omega_p$, with $\omega_p \sim J_c^{1/2}$. The fundamental
response frequency of Josephson devices, the Josephson frequency
$\omega_J$, also depends on the critical current density. On the other
hand, a drawback is that high critical current densities  lead to
large subgap leakage currents \cite{kleinsasser} and junction
characteristics degrade rapidly  with increasing $J_c$.

Variations in the critical current density also influence the $I-V$
 characteristics introducing steps under the influence of both a static
 bias current and the irradiation with microwaves \cite{reinisch}. In
 that case the variation is quite smooth (of $sech$ type ), so that
 the fluxon and its motion could be described by a small number of
 collective coordinates. Interesting behavior is also seen in both
 the static and dynamic properties for the case of a spatially
 modulated $J_c$ with the existence of "supersoliton" excitations
 \cite{oboznov,larsen} and the case of columnar         defects
 \cite{tinkham,balents} or disordered defects \cite{fehren}.

 The trapping of fluxons can be seen in the $I_{max}(H)$ curves where
 we also expect important hysteresis phenomena when scanning the
 external magnetic field. The hysteresis can be due to two reasons:
 (i) One is due to the non-monotonic relation between flux and
 external magnetic field \cite{caputo} arising from the induced
 internal currents, and (ii) from the trapping or detrapping of
 fluxons by defects.
 The effect of a defect on a fluxon and the
 strength of the depinning field depends strongly in the size of the
 defect, the type of defect and the position of the defect. Here we
 will consider case where the widths of the defects is of the order
 of the Josephson penetration depth. In this range we expect the
 strongest coupling between fluxons and defects. We will also consider
the case of a few defects in the low magnetic field region where
pinning and coercive effects are important.

 The organization of
 the paper is as follows. In Sec. 2 the sine-Gordon model for a
 Josephson junction is presented. In Sec. 3 we present the results
 of the critical current $I_{max}$ versus the magnetic field of a
 junction with an asymmetrically positioned  defect. The variation of
 the $I_{max}$ and the flux content  $N_f$ with the defect critical
 current density and the position are presented in sections 4 and 5
respectively. The effect of multiple pinning centers is examined in
sections 6 and 7. In Sec. 8 we examine a defect with a smooth
variation of the critical current density. In the last section we
summarize our results.

\section{The junction geometry} The
electrodynamics of a long Josephson junction is characterized from the
phase difference $\phi(x)$ of the order parameter in the two
superconducting regions. The spatial variation of $\phi(x)$ induces a
local magnetic field given by the expression
\begin{equation}
{\cal H}(x) =  \frac{d { \phi}(x)}{dx}.~~~\label{phix},
\end{equation}
in units of $H_0=\frac{\Phi_0}{2\pi d \lambda_J}$, where $\Phi_0$ is
the quantum of flux, $d$ is the magnetic thickness and $\lambda_J$ is
the Josephson penetration depth. The magnetic thickness is given
by $d=2\lambda_L +t$ where $\lambda_L$ is the London penetration
depth in the two superconductors and $t$ is the oxide layer thickness.
The $\lambda_J$ is also taken as the unit of length. The current
transport across the junction is taken to be along the $z$ direction.
We describe a 1-D junction with width $w$ (normalized to $\lambda_J$)
in the $y$ direction, small compared to unity. The normalized length
in the $x-$direction is  $\ell$. The superconducting phase difference
$\phi(x)$ across the defected junction is then the solution of the
sine-Gordon equation \begin{equation} \frac{d^2 {
\phi}(x)}{dx^2} = \widetilde{J}_c(x)\sin[{\phi(x)}],~~~\label{eq01}
\end{equation}
with the inline boundary condition \begin{equation}
\frac{d { \phi}}{dx}\left|_{x=\pm\frac{\ell}{2}}\right. =\pm
\frac{{I}}{2}+H,~~~\label{eq02} \end{equation} where $I$ and $H$ are
the normalized bias current and external magnetic field. $\widetilde
J_c(x)$ is the local critical current density which is $\widetilde
J_c=1$ in the homogeneous part of the junction and $\widetilde
J_c=j_d$ in the defect. Thus the spatially varying critical current
density is normalized to its value in the undefected part of the
junction $J_0$ and the $\lambda_J$ used above is given by
\[\lambda_J=\sqrt{\frac{\Phi_0}{2\pi \mu_0 d  J_0}},\] where $\mu_0$
is the free space magnetic permeability. One can also define a
spatially dependent  Josephson penetration depth  by introducing
$\widetilde{J}_c(x)$ instead of $J_0$. This is a more useful quantity
in the case of weak distributed defects.

In the case of overlap boundary conditions Eqs. (\ref{eq01}) and
(\ref{eq02}) are modified as
\begin{equation} \frac{d^2 {
\phi}(x)}{dx^2} = \widetilde{J}_c(x)\sin[{\phi(x)}] -I,~~~\label{eq0a}
\end{equation}
and \begin{equation}
\frac{d { \phi}}{dx}\left|_{x=\pm\frac{\ell}{2}}\right. =
H.~~~\label{eq0b} \end{equation}

We can classify the
different solutions obtained from Eq. (\ref{eq01}) with their magnetic
flux content \begin{equation}  N_f = \frac{1}{2 \pi}
(\phi_R-\phi_L) ,~~~\label{phi} \end{equation}
in units of $\Phi_0$, where $\phi_{R(L)}$ is
the value of $\phi(x)$ at the right(left) edge of the junction.
Knowing the magnetic flux one can also obtain  the magnetization from
\begin{equation} M= \frac{2\pi}{\ell} N_f-H. ~~~~~\label{magn}
\end{equation}
For the perfect junction, a quantity of interest is the critical
magnetic field for flux penetration from the edges, denoted by
$H_{c1}$. For a long junction it is equal to 2 while for a short it
depends on the junction length. Due to the existence of the defect
this value can be modified since we have the possibility of trapping
at the defects. For a short junction we have penetration of the
external field in the junction length, so that the magnetization
approaches zero. For a long junction it is a non-monotonic function of
the external field $H$.

To check the stability
we consider small perturbations $u(x,t)=v(x)e^{st}$ on the static
solution $\phi(x)$, and linearize the time-dependent sine-Gordon
equation to obtain:\begin{equation}   \frac{d^2 v(x)}{dx^2}
-\widetilde J_c(x)\cos\phi(x) v(x)= \lambda v(x) ,~~~\label{eq10}
\end{equation} under the boundary conditions
$$\frac{dv(x)}{dx}|_{x=\pm \frac{\ell}{2}}=0,$$ where $\lambda=-s^2$.
It is seen that if the eigenvalue equation has a negative eigenvalue
the static solution $\phi(x)$ is unstable. There is considerable
eigenvalue crossing so that we must monitor several low eigenvalues.
This is especially true near the onset of instabilities.

\section{Asymmetric defect} In the following
we will consider the variation of the maximum critical current as a
function of the magnetic field for several defect structures. We
start with a long ($L>\lambda_J$)  junction of normalized length
$\ell=10$  with a defect
of length $d=2$ which is placed $D=1.4$ from the right edge. Thus the
defect is of the order of $\lambda_J$. We plot in Fig. 2a the
maximum critical current $I_{max}$ variation with the magnetic field.
The different curves correspond to phase distributions for which we
have a maximum current at a given value of the magnetic field $H$. The
overlapping curves called modes have different flux content as seen in
Fig. 2b where we plot the magnetic flux in units of $\Phi_0$ for zero
current versus the external field. The magnetic flux is only a weak
function of the external current.

For the perfect junction there is no
overlap  in the magnetic flux between the different modes. In fact
each mode has flux content between $n\Phi_0$ and $(n+1)\Phi_0$ and
therefore is labelled the ($n,n+1$) mode \cite{caputo}. Here in the
case of the defect the range (at zero current) of flux for each mode
can be quite different and the labelling is with a single index
$n=0,1,2, ...$ corresponding in several cases to the $(0,1)$, ($1,2$),
($2,3$),... modes of the perfect junction. There are in several cases
several modes with similar flux. To distinguish them we add a letter
following the index $n$.

The maximum $I_{max}$ is obtained for mode $1$  and the increase
comes from the  trapping of flux by the defect. We have to note that
the $(d,e)$ part of this mode is a continuation of the $(a,b)$ part of
mode $0$. In both cases we have entrance of flux from the no defect
part of the junction and the instability in the critical current
occurs when $\phi(-\ell/2)=\pi$. Here and in the following  we will
take this to mean equal to $\pi$ modulo $2\pi$. For the maximum
current (at $H < 0$) the equation is $H-I/2=-2$. This can be
understood from the pendulum phase diagram, where the $\phi_x=-2$ is
the extremum slope, and thus the relation $I_{max}=4+2H$ holds. For
the $(b,c)$ part of mode $0$ the flux enters from the right where the
defect is. This reduces the critical current compared to the perfect
junction $0$ mode \cite{caputo,owen}. Note that $0$-mode has its
critical current $I_{max}$ peak slightly to the left of $H=0$ in the
$I_{max}$ vs $H$ diagram, and to the left of $N_f=0$ in an $I_{max}$
vs $N_f$ diagram (see Fig. 11a). Also in the absence of current,
reversing the  direction of $H$ only changes the sign of the slope
$d\phi/dx$, but the phase difference (in absolute value) at the two
ends will be the same. Thus the $0$ mode at $I=0$ extends for $-1.6\le
H \le 1.6$. This is not clearly seen due to curve overlapping in the
left side. Comparing the $I_{max}$ for the modes $1$ and $-1$ we see
that $I_{max}(1)>I_{max}(-1)$. In both cases a fluxon (or antifluxon)
is trapped in the defect. The major difference in the $I_{max}$ comes
mainly from the phase distribution which in the mode $1$ case leads to
a large positive net current in the undefected side, while in the $-1$
mode the net current in the undefected side is very small.

For the mode $1$, at $H\approx 0$ and
zero current the instability happens due to the competition of the
slope of the phase at the defect center and at the right edge, while
at the other end at $H=2$, the field at the defect center becomes
equal to the external field applied at the boundaries and there is no
such competition. In this case the instability sets in due to the
critical value of the phase at the undefected boundary (i.e.
 $\phi_x(-\ell/2)=2$). The situation is analogous for
the mode $-1$. For $H\approx 0$ the instability sets in due to the
depinning of the antifluxon while for $H=-2$ due to the critical
value of the phase at the free defect part of the junction.
For the mode $0$ we have no fluxon trapping at the defect,
even though the instability at the two extremes with $H=\pm1.6$ 
at zero current is
caused from the tendency to trap a fluxon or antifluxon
correspondingly at the defect.
 At higher
values of the magnetic field ($|H|>1.6$)  we have stability
for a range of non-vanishing current values as will be discussed
below.   Thus this value can be considered as the minimum value for
the introduction of fluxons in the junction. Let as remark that for
the perfect junction, or a junction with a centered defect, the
corresponding values for fluxon introduction would be equal to $2$.
Thus there is a decrease of the critical field as the defect moves
away from the center. For the $0$ mode a centered defect would have
no influence on the solution.

The results for the maximum current are in agreement with
the stability analysis. In Fig. 2c we present the lowest eigenvalue
$\lambda_1$ for the different modes in zero external current $I=0$ as
a function of the magnetic field $H$. The sudden change in slope 
for the modes $-1, 1$ is
because at that point a new eigenvalue becomes lower. The $\lambda_1$
is positive denoting stability and becomes zero at the critical value
of the magnetic field, where a mode terminates. The symmetry about
zero magnetic field is due to the symmetric boundary conditions for
$I=0$.  Change of the sign of $H$ changes the sign of the phase
distribution, but the $\cos\phi$ in (\ref{eq10}) remains unchanged.
This symmetry is being lost when a finite current is also applied.
Also there are solutions (not presented in the figure) for which the
stability analysis gives negative eigenvalues i.e instability. These
solutions may be stabilized when we insert multiple impurities.

  In Fig. 3a
we specifically draw only the $1$ mode, to be discussed in more
detail. Here we changed the procedure, in searching for the maximum
current. Up to now we followed the standard experimental procedure,
i.e. we scan the magnetic field and for each value of $H$ we increase
the current $I$, starting from $I=0$, until we reach the maximum
current. Here we consider the possibility that for $I>0$ there is
also a lower bound in the value of the current for some values of the
magnetic field. This requires a search where we vary both $H$ and
$I$ simultaneously. Thus we see that for $H<0$ there is a lower
bound given approximately by the line $H+I/2\approx H_{cl} $, where
$H_{cl}\approx 0$ is the critical value of $H$ at $I=0$, for which we
 have depinning of the trapped fluxon. Over this curve the slope
 $\phi_x$ at the right end (near defect) is kept constant and equal to
 $H_{cl}$ and above this line the fluxon remains pinned and it should
 be stable. This line ends at $H=-1$, since in that case the extremum
 value $\phi_x=-2$ is reached in the left end. Increasing now in that
 range of $H$ the bias current we find that also the $I_{max}$ curve
 extends further to the left. The equation for this line is
 approximately given by $H-I/2=-2$, with an extremum at
 $\phi_x(-\ell/2)=-2$. Thus the instability on this line arises from
 the left side (far from the defect). It extends up to $H=-1$ for a
 long junction and joins the other line $H+I/2=H_{cl}$.

 The above calculations were done for a
 long junction so that the fields at the two ends do not interfere.
 For shorter length however the two ends feel each other and in that
 case the two instabilities are not independent. This means that the
 tail of the defect free side field will compete with the slope of
 the trapped field. Then the two lines $H+I/2=H_{cl}$, and
 $H-I/2=-2$, end  before they meet (at $H\approx-1$) at a cutoff
 magnetic field. Also for short junctions we expect the straight
 lines to have some curvature.
 A similar discussion holds for the
 right end of the $1$ mode. Again there is a lower current
 (positive) bound given by $H-I/2=2$ due to instability at the left end
 ($\phi_x(-\frac{\ell}{2})=2$), and an upper bound given by
 $H+I/2=H_{cr}$, where $H_{cr}\approx 2.8$, due to fluxon depinning.
 On the same diagram, we show the lower bound for negative currents.
 Thus we see that there is strong asymmetry for positive and negative
 currents. Remark that for negative currents the mode $1$ is very
 similar to the mode (1,2) with no defect \cite{caputo}. This is
 because the right boundary is determined by an instability at the
 undefected side. The left boundary is again very close because
 $H_{cl}\approx 0$. So an interesting effect of the defect is that we
 have this strong asymmetry for positive and negative currents.

 We would get a similar picture if we considered
 the $-1$ mode. In fact we get the same curves (as for mode $1$) if we
 put $I\rightarrow -I$ and $H\rightarrow -H$. This is consistent with
 the $-1$ mode shown in Fig. 2a. The  discussion can also be
 extended to the other modes. In Fig. 3b we show the result of a
 similar scan for the $0$ mode, but for the sake of shortness we will
 not discuss the $-1,-2,2$ modes. In any case when the number of
 fluxons increases one must rely on numerical calculations rather than
 simple arguments.

 \section{Variation with the defect critical current}
 In the previous section we considered the case of a microresistance
 defect. With present day masking  techniques we
 can also consider any finite critical current (lower or higher) in
 the defect. This situation can also arise very often in junctions
 with high critical current densities, where small variations in the
 thickness can create strong critical current density variations. Thus
 for the previous asymmetric defect configurations we will study the
 effect of the defect critical current density in the magnetic
 interference pattern $I_{max}(H)$. We will concentrate on the $0$ and
 $1$ modes.

 {\it (i) mode $0$:}

 In Fig. 4  we see the $I_{max}(H)$ variation
 for the mode $0$ for decreasing values of the critical current
 density from $j_d=2$, to $j_d=0$. Let us discuss first the case
 for $j_d \le1$. For the perfect junction where $j_d=1$ we have a
 symmetric distribution about $H=0$. As we decrease $j_d$ the flux
 content of this mode (and the extremum $H$) is symmetrically reduced
 (see Fig. 2b). It is not apparent from the drawing, due to the
 superposition of several curves on the left side of the diagram, but
 as expected the range of the magnetic field is symmetric about $H=0$
 at zero current. The corresponding $I_{max}(H)$ curves, however, are
 not symmetric. The right hand side of the curves is displaced towards
 smaller critical fields with decreasing $j_d$.  This means that the
 critical field at $I=0$ to introduce a fluxon from the ends is
 decreased due to the existence of the defect which acts with an
 attractive force on the fluxon. The curves are linear and can be
 approximated by the equation $I(H)=4-2(H+\delta H_c)$, where $\delta
 H_c$ is the decrease in the critical field $H_{c0}$ and depends on
 $j_d$. A similar decrease happens for negative magnetic fields where
 the defect tries to pin an antifluxon. Even for higher currents the
 right side critical field is determined by the tendency of the defect
 to attract a fluxon.  The left hand side, however remains rigid (but
 is shifted along the line). This is due to the entrance of magnetic
 flux from that part of the junction where there is no defect. The
 instability in the critical current occurs when ($\phi(-\ell/2)=\pi$)
 for every value of $j_d$. From the pendulum phase diagram which is
 the classical analog of the Josephson junction, the extremum occurs
 at $\partial_x\phi (-\frac{\ell}{2})=-2$, or $H-I/2=-2$ which is the
 equation for this triangular side. At near zero current the critical
 field is influenced from the attractive action of the defect. At low
 currents and extreme negative magnetic field the $I_{max}$ curve
 shows a re-entrance behavior so that it is not stable at low and high
 currents, but only for a finite intermediate range of current values.
 This way we reconcile the different origins of the instability
 mechanisms $\phi(-l/2)=\pi$ at high current and the defect influence
 discussed for the right hand side of the mode.

For $j_d>1$ we see an increase in the $I_{max}$, while  the critical
magnetic field at $I=0$ remains almost constant at about
$H_{cr}\approx 1.9$. The instability at that point is due to the
trapping of flux in the region between the positive defect and the
right edge of the junction. The field for that is expected to be near
$H=2$ if the right undefected part is of length of the order of
Josephson length. Thus it is the same value for flux penetration from
the perfect junction edges. It will vary weakly with $j_d$.

 {\it (ii) mode $1$:}

 The mode $1$ in the perfect
 junction has a full fluxon for magnetic field $H=0.07$. The phase
 distribution is about $\theta=\pi$ where the energy has a minimum.
 At the end of this mode at $H=2.07$ where two fluxons have entered
 the junction the phase changes from $\phi(-\ell/2)=-\pi$ to
 $\phi(\ell/2)=3\pi$. When the defect is inserted this mode is
 significantly modified due to the flux trapping in the defect.

 In Fig. 5 we see the magnetic interference pattern for this mode for
 different values of the defect critical current density.
 For $0 < j_d < 0.7$ the $I_{max}$ vs $H$ curves are displaced
 downwards, and a fluxon is trapped in the defect. We notice that all
 the curves for $j_d<0.7$ have the same critical magnetic field
 $H=2$ for $I=0$. This is because at this end of the mode, at $I=0$
 the instability arises at the side with no defect where the phase
 reaches the critical value $\phi=\pi$ (modulo $2\pi$). Of course as
 discussed in the previous section we have a reentrant behavior
 above $H=2$. At the other end for small magnetic field the instability
 is due to depinning of the trapped fluxon.
  For $0.7 < j_d \le 1.0$ the defect can trap the flux only for
 $H<H_{cd}$, where the value of the $H_{cd}$ depends on the defect
 critical current $j_d$ and in Fig. 5 it is shown for $j_d=0.9$.
 Notice that for this value of $j_d$ the fluxon is very weakly
 trapped, and the untrapping process happens slowly over a range of
 magnetic field values. For $H > H_{cd}$ the fluxon has moved away from
 the defect, and for this weak defect the junction does not feel it.
 The critical current goes abruptly close to the curve for the
 perfect junction. We conclude that the behavior of the junction for
 values of $j_d$ close to $j_d=1$ is determined by the ability of the
 defect to trap one fluxon. This can be seen also from the change in
 the lowest eigenvalue variation with the external field $H$, at
 values of the critical density $j_d > 0.7$, in Fig. 6.

For $j_d>1$ (thin lines in Fig. 5) it has a similar form as for
$j_d=1$, i.e. there is no fluxon trapping. Again, as in the $0$ mode,
the $H_{cr}$  at $I=0$ stays around $2.0$ and is again due to the
trapping of flux in the right edge.

 In Fig. 7 we present the evolution of $I_{max}$  with
 the defect critical current density $j_d$ for a magnetic field
 $H=1.5$. For this value of the magnetic field there are no solutions
 with trapped fluxons for $j_d > 0.83$. The lowest eigenvalue at
 $I=0$ becomes zero at this point. For $j_d > 0.83$ and $H > 1.5$ there
 are solutions which are not trapped. For these solutions the maximum
 current coincides with the one of the perfect
 junction and there is a discontinuity in the curves. Notice the point
 at $j_d=1.0$. In the same figure we also show the magnetic flux at
 $I=0$ and at $I_{max}$ which is almost constant as a function 
 of $j_d$ as expected, with
 small difference between the two different current curves.

 \section{Variation with the defect position} In Fig. 8a
 we see the evolution of the critical current at zero magnetic field
 as we move the defect from the right edge of the junction $D=0$ to
 the left edge where $D=8$. The position is measured from the edge of
 the junction to the nearest edge of the defect. We examine the
 several modes separately:

 {\it (i) mode $0$}

 For this mode and for
 $I=0$, we are able to find solutions for all the defect positions.
 As we can see in Fig. 8b the corresponding magnetic flux at $I_{max}$
 is slowly changing and equal to zero when the defect is in the
 junction center.
 But when the defect is placed close to the ends the magnetic flux at
 the maximum current deviates from zero. The
 critical current for this mode is symmetric for defect
 positions about the junction center, and has its maximum value when
 the defect is at the center. This is because at that position it
 does not influence the solution at the edges which is very close to
 the undefected case, while near the center the phase is almost
 zero. But when the defect comes close to the junction ends the
 defect cuts into the area by which the current flows, and the
 critical current is reduced.

 For even smaller distances $D=0.2$, and
 $D=0$ there is a jump to  solutions which correspond to a
 current, which is much higher than that of the $0$ mode for
 nearby $D$ values. This is because the defect cuts negative current
 regions  and for this position we have an increase of the critical
 current. In fact these solutions (see $++$ symbols in Fig. 8a) are
 very close to the solutions of a perfect junction within the
 undefected area, except that now the defect at the edge can give
 contribution to the flux but no contribution to the current. Thus the
 flux is much higher than that of the $0$ mode and it approaches that
 of mode $1$.  Nevertheless these points should be considered as a
 separate mode. In fact they are part of a branch (crosses). In these
 distances there are no other modes for $H=0$. Similar results were
 obtained by Chow $et$ $al.$ \cite{chow} where they attributed this
 enhancement in the $I_{max}$ for small distances to a self field
 which was generated by the current, penetrating into the defect and
 resisted any further penetration of field. To overcome this
 resistance it was necessary to apply a higher current. But they do
 not distinguish between modes with different flux content, and their
 evolution with the defect position.

 {\it (ii) modes $1$, $-1$}

 For these modes we do not have solutions for all the defect positions
 at $I=0$ and $H=0$, but only in the range $1.4 < D < 6.6$ as seen in
 Fig. 8c where the lowest eigenvalue is plotted as a function of the
 defect position for the different modes. The curves for the 1 and -1
 modes coincide, while the 0 mode shows a change of slope
 corresponding to the last two points ( $D=0 $ and 0.2 discussed above)
 which belong to another curve.  Mode $-1$ has a trapped antifluxon in
 the defect. When the defect is to the left ($4.0 < D <6.5$), then the
 instability in the current of mode $-1$ at $H=0$ is created at the
 right-end of the junction when the phase reaches the critical value
 $\phi(l/2)=\pi$.  This instability occurs for currents which are less
 than those necessary to unpin the antifluxon. Notice that increasing
 the current there is no competition with the slope of the antifluxon
 trapped in the left end. Thus at this point  (for $ 4.0 < D < 6.5 $)
 the maximum current is very close to the undefected junction mode
 $0$, except that in this case $N_f\approx -1$ is close to an
 antifluxon.  At the other end ($D < 4.0$) the instability for mode
 $-1$ is caused by the depinning action of the applied current, which
 takes now much smaller values (close to zero) because of competition
 with the pinned fluxon.  The phase distribution at the defect free
 end is that expected for $H=0$ and $I$ close to zero. The mode $1$
 with a fluxon trapped has a symmetrically reflected (about the
 center)  form in $I_{max}$ vs $D$ and the instability for $D>4.0$
 occurs at the left end of the junction, which is the opposite case of
 mode $-1$. The eigenvalue becomes zero at the positions $D=1.4$ and
 $D=6.6$. The $\lambda_1(D)$ curve coincides for the modes $1$ and
 $-1$ due to the fact that the phase distributions for the same $D$
 for these modes are symmetric about $x=L/2$, and the $\cos\phi(x)$
 that enters the eigenvalue equation is the same.

 In the rest we
 examine the variation of the critical value at which the instability
 sets in, as we scan the magnetic field in the positive (negative)
 direction $H_{cr} (H_{cl})$ for zero current, for the different
 modes, as a function of defect position. This instability can be
 attributed to the pinning, or the depinning field or to the
 critical value of $\frac{d\phi}{dx}$ at the defect free  edge,
depending on the particular mode that we are considering.
Explicitly for the mode $0$
the instability in the $H_{cl} (H_{cr})$ is due to the pinning of a
fluxon (antifluxon), respectively. In this mode the defect has no
influence for positions close to the center as seen in Fig. 9a and 9b.
However as we move the defect close to the edges the pinning field
$H_{cl} (H_{cr})$ is reduced in absolute value because it is easier to
trap  a fluxon (antifluxon).
For the mode $1$ the $H_{cr}$ is constant for
all defect positions. This is due to the fact that at $I=0$ it is
the phase distribution at the undefected  edge of the junction that
determines the instability. Notice that due to the reentrant character
the critical  magnetic field takes higher values at larger
bias currents which vary with defect position. The $H_{cl}$ curve
depends on the phase distribution near the defect and therefore is
strongly defect position dependent.
For the mode $-1$ the
picture is reversed compared with the $1$ mode. In this case the
$H_{cl}$ is constant while the $H_{cr}$ varies with position. Note
that in this mode the depinning of an antifluxon is the reason that
causes the instability at $H_{cr}$.

\section{Two symmetric pinning centers}
As noted  defects (with $j_d<1$) or inhomogeneities
in the junction can play the role of pinning centers for a fluxon. In
this section we discuss more precisely the effect of multiple pinning
centers on the magnetic interference patterns $I_{max}(H)$ and the
flux distribution. The pinning effect of the Josephson junction has
also been analyzed in \cite{yamashita1,yamashita2}, by using a simple
mechanical analog. The analogies of the mixed state of type II
superconductors and vortex state of the Josephson junction has been
discussed in these references. In Fig 10a we present, as an example,
the critical current $I_{max}$ versus the magnetic field for a
junction which contains two defects of length $d=2$ placed
symmetrically at a distance $D=2$ from the junction's edges. We
 examine the following modes grouped according to flux content:

{\it (i) modes $0$, $0a$}

 These modes have magnetic flux antisymmetrical around zero field, as
 seen form Fig. 10b where the magnetic flux is plotted versus the
 magnetic field. At $I=0$ and magnetic field $H=-0.7$, the $0a$ mode
 contains one fluxon trapped in the left defect, while an antifluxon
 exists at the other part of the junction. 
 As $H$ increases towards $0.7$ the picture changes
 slowly, so that the antifluxon is pinned in the right defect. The
 stability analysis shows that this mode is unstable. We remark that
 there are also other unstable modes near zero flux, which we will
 not present here. For example there is another unstable mode with
 the same flux as $0a$ but a much higher critical current (the same as
 the $0$ mode). Mode $0$ has phase distributions which are similar to
 the corresponding mode of the homogeneous junction since it has no
 trapped flux in each defect.

 {\it (ii) modes $1l$, $1r$}

 These modes have
 magnetic flux close to unity, and are both stable. For the mode $1r$
 one fluxon has been trapped to the right defect, and in the mode
 $1l$, the vortex is trapped in the left defect. Due to the symmetry
 this mode has the same magnetization as the mode $1r$, but the
 critical current is reduced. The phase distribution for the modes
 $1r$, and $1l$, at zero current are related by
 $\phi_{1l}(x)=2\pi-\phi_{1r}(-x)$. The maximum field $H=1.9$ (at $I=0$)
 for both modes is determined by an instability at the defect free
 side.  At the other extreme
 there is a competition at the fluxon side between the applied field
 and the field created by the pinned fluxon. Thus the critical field
 at $H=-0.62$ can be considered as a coercive field and below this
 value the fluxon gets unpinned. The two modes have characteristically
 different currents and this depends on the current through the
 fluxon free defect, since the pinned fluxon itself gives no major
 contribution. Thus the maximum current is much larger for the $1r$
 mode. The opposite would be true if we look for negative currents.
There are also the symmetrically situated modes that correspond to
an antifluxon in the left or right defect, which are not shown in
 Fig. 10a. The respective flux is antisymmetric with $H$ around
 $H=0$.

 In Fig. 10a we also show the mode $2$ with flux around $2$
 fluxons. Several unstable modes are not shown, for the sake of
 clarity. Their analysis however, can show the connection between
 different modes, while a defect in the correct place with proper
 characteristics can stabilize these solutions.
 We conclude that
 depending on the positions where the vortex is trapped we may have
 modes with the same magnetic flux content, but different critical
 currents. Also due to soliton localization on the defects, we may
 have stable states with magnetic flux close to unity, for zero
 magnetic field. These states together with the one existing in the
 homogeneous junction form a collection of stable states in a large
 $H$ interval. We must comment here that states with unit flux, for
 zero magnetic field ($H=0$) exist in the homogeneous junction, as a
 continuation of the stable ($1$) mode to negative magnetic fields,
 but as we found in a previous work \cite{caputo}, are unstable. So
 we may argue here that the presence of defects stabilizes these
 states.

 In comparing the results for one (Fig. 2a) and two defects
 (Fig. 10a) we see some similarities and differences. In the case of
 two defects new modes appear but also the region of stability of the
 equivalent modes is different. This is more clearly seen in Fig. 11
 where we plot the $I_{max}$ vs $N_f$ for both cases.  This
 presentation is useful since the $N_f$ is a nonlinear function of
 $H$. This plot (Fig. 11a) is a combination of Figs. 2a and 2b.  We
 should point out that the maximum peak in the current in both case
 comes due to the trapping of a fluxon in the defect at the right
 side. The maximum of $0$-mode is very close in both cases and this
 happens because this mode does not involve fluxon trapping. The $1r$
 mode for the two defect case is very close to the $1$ mode of the
 single defect, since in both cases there is a fluxon trapped in the
 same side. In the two defect case we see an enlargement of the region
 of stability so that the modes overlap. The thin continuation lines
in modes $0$ and $1$ for the single defect are in the reentrant region
of flux as discussed in section 2.

 \section{Symmetric distribution of pinning centers}
 In this section  we study as an example the case where a junction of
 length $\ell=14.2$ contains three defects of length $d=2$, and the
 distance between them is $2$. The length was augmented, so that we
 keep the same width of the defects when we increase the number of the
 defects, since we saw that the width of the order $d =2$, gives the
 possibility of fluxon trapping and increased maximum current when the
 defect is situated asymmetrically. We will study the phase
 distribution at $I=0$ and try to extract information about the
 critical field values and magnetization. We find the following modes
 grouped according to flux content:

 {\it i) modes $0$, $0l$, $0r$, $0c$}

 In Fig. 12a we
 present the critical current versus the magnetic field for the modes
 with magnetic flux around  zero (see Fig. 12d). This is indicated by
 the $0$ symbol. There are four modes belonging in this category,
 which are stable. The solutions for the mode $0$ are similar to the
 homogeneous junction mode $0$, with no flux trapping in the defects.
 The only difference is that the instability in the critical field
 occurs when the phase at one edge, reaches a value, which is smaller
 (due to pinning) than the corresponding value for the undefected
 junction, which is $\phi(-\ell/2)< \pi$. The same was true for the
 two defect case. Mode $0c$ has the maximum critical current
 $I_{max}=5.08$ for $H=0$. One antifluxon is trapped to the leftmost
 defect, one fluxon to the rightmost, and the phase in the center
 defect is constant. The trapping at the edge defects leads to a
 positive current distribution between them, for this particular
 length, and enlarges the maximum current. The same type of mode was
 not found for the two defect case (with a shorter junction length),
 and we conclude that the extra defect along with the increased
 junction length stabilizes this solution. For the mode $0l$ one
 fluxon is trapped in the left defect where the phase changes about
 the value $\phi=\pi$. The antifluxon is distributed at the other two
 defects, where the phase is about the values $3\pi/2$ (or $\pi/2$),
 and we have a cancellation of the positive and negative current
 density in this region. Similar for the mode $0r$ the fluxon is
 trapped to the right defect, and the current is distributed with
 opposite sign to the other two defects.  These modes are similar to
 the $0a$ mode for the two defect case.

 {\it ii) modes $1l$, $1c$, $1r$}

 In Fig. 12b we see the maximum
 current versus the magnetic field for the modes with magnetic flux
 around $N_f=1$ (see Fig. 12e). There are three modes with flux close
 to $N_f=1$ each of which corresponds to the trapping of one fluxon in
 one defect. In the mode $1c$ the fluxon is trapped in the center
 defect. In the mode $1l$ ($1r$) it has been trapped in left (right)
 defect. Due to the symmetry the lowest eigenvalue, and the magnetic
 flux coincides for these two modes, but as we showed in the previous
 section, their critical currents are different, depending on the
 tunneling current distribution in the region with no trapping. The
$1r$ mode corresponds to a higher critical current.

 {\it iii) modes $2$, $2a$, $2b$}

 In Fig. 12c we see the maximum
 current versus the magnetic field for the modes with magnetic flux
 around  $N_f=2$ (see Fig. 12f). Only the mode $2$ corresponds to
 stable solutions. There we have two fluxons trapped in the side
 defects. In mode $2a$ one fluxon is trapped in the right defect,
 while in mode $2b$ this trapping occurs in the center defect. We
 conclude that distributed pinning centers are more effective in
 trapping the vortex, and lead to an increased critical current. Some
 conclusions  will continue to be valid for larger number of defects
 where we keep the defect width and separation fixed. In that case we
 also expect the results to change significantly when there is either
 a periodic array of defects, where we expect higher fluxon modes to
 give the highest current peak \cite{tinkham}.

 \section{Defect with a smooth variation of current density}
 Up to now we considered
 defects with abrupt changes in the local critical current density 
 and the question
 arises whether the abruptness of $J_c$ variation is crucial in the
 significant change in $I_{max}$ for the $n=1$ mode. We will see that
 similar effects exist for smoother variation, where again the fluxon
 pinning is an important feature. For this reason we chose a single
 defect at the junction center with a smoothly varying critical current
 density
 given by \begin{equation}\widetilde J_c (x)=\tanh ^2\left[
 \frac{2}{\mu}(x-x_0) \right],~~~\label{delta}\end{equation}
 where the defect is centered at $x_0$, and the width is
 determined by  $\mu$.
 In Fig. 13 we show the results for the case $x_0=7.6$ and $\mu=2$,
 which can be compared with the results of the asymmetric defect in
 Fig. 2a. For the modes shown the curves are very similar and thus we
 see that the main results survive since the defect strengths are
 similar. Of course  there is a quantitative difference. But most of
 the stability criteria described earlier are still valid.

 In Fig. 14 we consider the effect of the form of current input and
 compare the case of inline with overlap for a smooth defect situated
 at the center of the junction i.e. $x_0=0$ with $\mu=0.5$. In Fig.
 14a we present the $I_{max}$ for inline boundary conditions, and we
 show only the $-1,0,1$ modes. The $0$ mode is not influenced at all
 from the defect since all the phase variation is at the boundaries.
 There is a strong similarity with $I_{max}$ for  the $1$, and $-1$
 modes. The reason is that in these cases there is a trapped fluxon
 or antifluxon at the center and at zero current and magnetic field
 the phase variation dies out at the boundaries. Thus when increasing
 the current at $H=0$ towards $I_{max}$ we have the same situation at
 the boundaries as for the $0$ mode and  the instability happens at
 close $I_{max}$ values.  Of course due to the pinning, the fluxon
 content is very different from  the $0$ mode. The $-1$ and $1$ modes
 have an enhanced $I_{max}$ and the  small difference in $I_{max}$
 from the $0$ mode is attributed to the small influence of the trapped
 fluxon to the boundaries. Let us remark that a similar situation was
 seen in Fig.  8a for the square well defect, when the defect position
 is at the center for $H=0$. By comparing with Fig. 9 the $H_{cl}$
 and $H_{cr}$ values we see close agreement with the case of $j_d=0$
 in the defect.  These results could change for a smaller length
 junction or if we move the defect towards the edges (as seen in Fig.
 8a).

 For the same defect we also investigated the effect of
 the overlap current input, where the current is distributed along
 the whole junction. In Fig. 14(b) we present the maximum current per
 unit junction length versus the magnetic field, and it should
 be compared to the inline  case in Fig. 14(a). We see a significant change
 for the $-1$ and $1$ modes. Of course at $I=0$ both current
 inputs give the same solutions, but $I_{max}$ is much smaller for the
 overlap boundary conditions. This is from the fact that due to the
 applied current the fluxon is pushed against the pinning barrier
 until it is overcome at the critical current. In the absence of
 applied current the phase at the defect center is $\phi(0)=\pi$,
 while the application of the current pushes the fluxon to the edge of
 the defect which is taken to be near the point where the curvature of
 the defect critical current distribution changes sign. So we can
 consider in this case this maximum current as a measure of the
 pinning force.

 In Fig. 15a we plot the magnetic flux $N_f$ at zero
 current versus the magnetic field $H$ for the  inline  case.
 The lowest eigenvalues for the different modes versus the magnetic
 field are seen in Fig. 15b. For a homogeneous junction the $0$ mode
 is the only stable state available at $H=0$.  However in the problem
 we consider here, the mode $1$($-1$) exists and it is stable for
 $H=0$ and corresponds to the localization of the soliton
 (antisoliton) in the inhomogeneity. For these modes we have pinned
 flux at $H=0$, with $\phi(0)=\pi$, and $\frac{d\phi}{dx}=2$. In Fig.
 16 we show the evolution of $\phi$ and $\frac{d\phi}{dx}$, for the
 mode $1$, as we change the magnetic field at $I=0$. Near $H=-1.9$ the
 fluxon content is near zero and for $H<-1.9$ an instability sets in
 due to the depinning of the fluxon. This is because the slope at the
 pinned fluxon competes with the opposite slope tried to be imposed by
 the external negative magnetic field at the boundaries. At the other
 end the flux is equal to $2$, and the instability sets in when $\phi$
 at the boundaries approaches $\pi$ (or odd multiplies). The range of
 $H$ values for the $1$ mode, when the defect is at the center is
 significantly broadened and gives a corresponding range for the flux
 of two fluxons. Usually each mode has about one extra fluxon and in
 particular for the perfect junction it contains only one extra
 fluxon. This is because the defect is at the center and far enough
 from the edges where the magnetic field is applied and therefore
 even for negative fields there is no significant competition, with
 the field at the defect center. This is especially true when the
 distance of the defect from the edges is grater than
 $2\lambda_J$. When however the defect is near the edge the
 instability sets in before
we cross to negative magnetic fields.

The maximum current $I_{max}$ for the mode $0$ is greater than in the
modes $1$,$-1$, but is reduced compared with the $I_{max}$ for the
mode $0$ in the homogeneous junction, in zero field. In
\cite{filippov} they approximated this reduction in an analytical
calculation using a delta function for the defect potential, and they
found $\Delta I_{max}=-\mu/2L\approx0.02$. In \cite{filippov} they
arrived at the analytical result, by minimizing the fluxon free
energy, for the maximum overlap current versus the magnetic field
$H$, for these modes, which is a good approximation of the numerical
solution we consider here in the limit $L \gg 1$.

\section{Conclusions}  In several applications it is
desirable to work in an extremum of the current for a region of the
magnetic field. This can be achieved by the appropriate distribution
of defects so that the negative lobes of the current distribution in
the junction due to the fluxons are trapped in the defect with no
contribution to the current. Of course if the defect is isolated
(far from other defects or the edges) we expect zero contribution to
the current. Due to the effect of the applied current and magnetic
field at the boundaries in certain cases we can obtain  positive
current lobes outside the defect. In several cases in section 3 this
was the reason for the  increased current. Because the control of
magnetic field is very easy compared to other system parameters (like
temperature, disorder, etc.) the measurement of the effect of magnetic
field on junction behavior, provides a convenient probe for the
junction. The calculation of the $I_{max}$ can characterize the
quality of the junction or verify the assumed distribution of defects
when they are artificially produced. The spatial variation of the
critical current density on low $T_c$-layered junctions, and high
$T_c$ grain boundary junctions can be directly imaged with a spatial
resolution of $1\mu m$ using low temperature scanning electron
microscopy $(LTSEM)$ \cite{huebener,gerdemann}. Information on smaller
scale inhomogeneities has to rely on the magnetic field dependence of
the maximum tunneling current $I_{max}$.

The purpose of this paper is the consideration of large defects in
order to study the interaction between fluxons and defects and give
estimates of the coercive field for pinning or depinning of a
fluxon from a defect. The region of consideration puts us far from
the region of perturbation calculations and is amenable to direct
experimental verification since it is easy to design a junction
with the above characteristics. The defects influence strongly the low
fluxon modes. At high magnetic fields larger than the depinning field
of a single fluxon we expect only minor effect and fluxon trapping. Of
course for a large number of defects interesting behavior can be
obtained.\cite{oboznov,larsen} The interaction between fluxons in the
few defect case also assists to overcome coercive fields and untrap
fluxons. The results of two trapped fluxons in the two defect case
show that the fluxons are strongly coupled and one cannot consider an
exponential interaction type potential between the fluxons. Also the
critical current in a long junction, cannot be calculated as the
Fourier transform of the spatial distribution of the critical current
density $J_c(x)$, at least for weak magnetic fields.  For strong
magnetic fields, where we have the field penetrating uniformly the
junction, as is the case for short junctions, we recover the
diffraction like pattern.

In summary we saw that the bounds of the
different modes determined by the stability analysis depend on two
factors: (i) the instability at the boundaries away from the defect
when $\phi_x$ reaches its extremal values equal to $\pm 2$, and (ii)
the instability due to the pinning or depinning of a fluxon by the
defect. If the junction is near one end then we saw that both criteria
play a role in determining the instability, independently in different
areas. In general, however, there will be coupling between defects and
the edges (surface  defects) especially in the case of multiple
defects.  Defects also introduce hysteresis phenomena which are weaker
in the case of smooth defects. We also saw that due to fluxon trapping
in general we see a reentrant behavior, i.e. there are regions of
magnetic field for which there is both an upper and a lower bound on
the maximum current.
We also find  that
due to the pinning of magnetic flux from the defect there exist
additional stable states in a large interval of the magnetic
field. The abrupt change in the critical current density is not
crucial for the trapping. Similar results are expected from smooth
defects, with quantitative differences. The above results can be
checked experimentally since it is easy to design a junction with a
particular defect structure, using masking techniques.
In fact a few parameters or characteristics could give at least
partial information on defect properties. In particular the measurement
of $H_{cr} $ or $H_{cl}$ can give some information of the defects near
the edges. Also one can imagine the situation where we scan locally
with an electron beam affecting thus the local critical current and
observe the variation of the $I_{max} $ as we increase the heating.
Once a fluxon is trapped we can decrease the heating (or increase
$j_d$) and observe the variation of $I_{max}$. Thus one can have
pieces of information to put together in guessing the defect structure
that might fit the whole $I_{max}$ pattern. The extension to many
  defects requires considerable numerical work. It is hoped, however,
  that some of the stability criteria will still be useful.

{\bf Acknowledgements} One of us (N. S.) would like to acknowledge the
ESF/FERLIN programme for partial support. Part of this work was done
under grant PENED 2028 of the Greek Secretariat for Science
and Research.

\bibliographystyle{prsty}
\newpage

 \begin{figure}
 \centerline{\psfig{figure=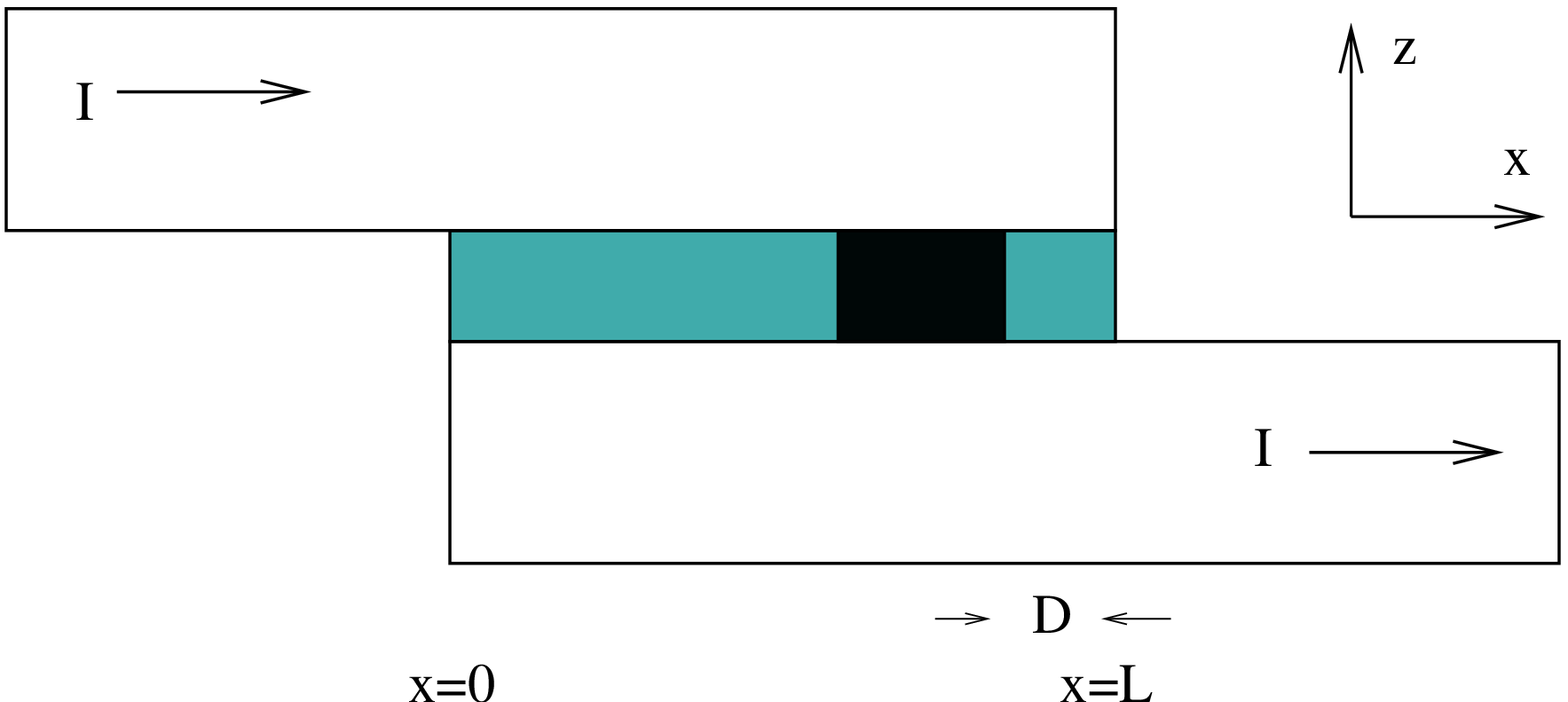,width=8.5cm,angle=0}} 
 \caption{The geometry of the junction. The dark shaded
 region marks the defect in the intermediate
 layer. $\ell$ is the junction length and $D$
 the separation between the left edges of the
 defect and the junction. }\label{fig1}\end{figure}

 \begin{figure}
 \centerline{\psfig{figure=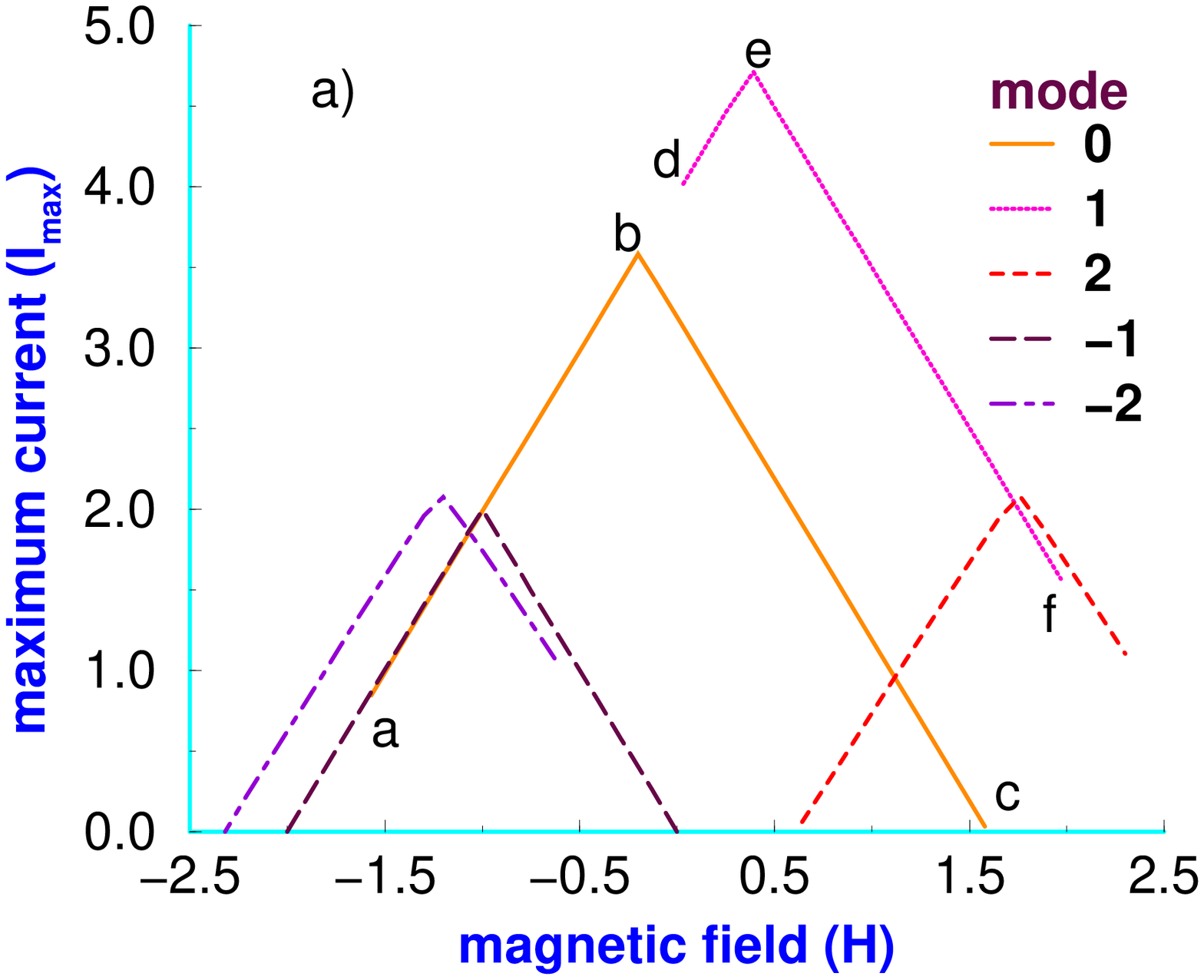,width=8.5cm,angle=0}}
 \centerline{\psfig{figure=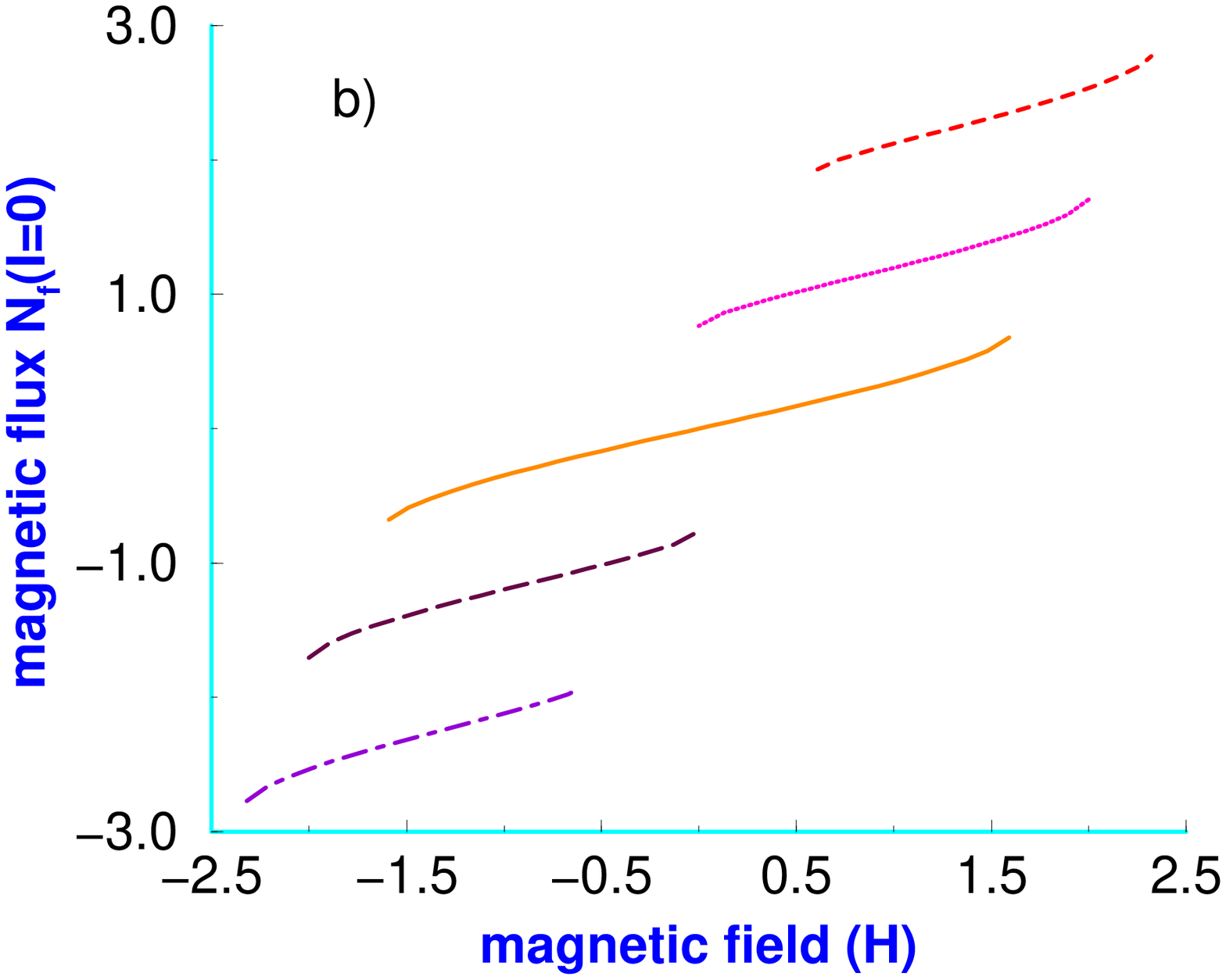,width=8.5cm,angle=0}}
 \centerline{\psfig{figure=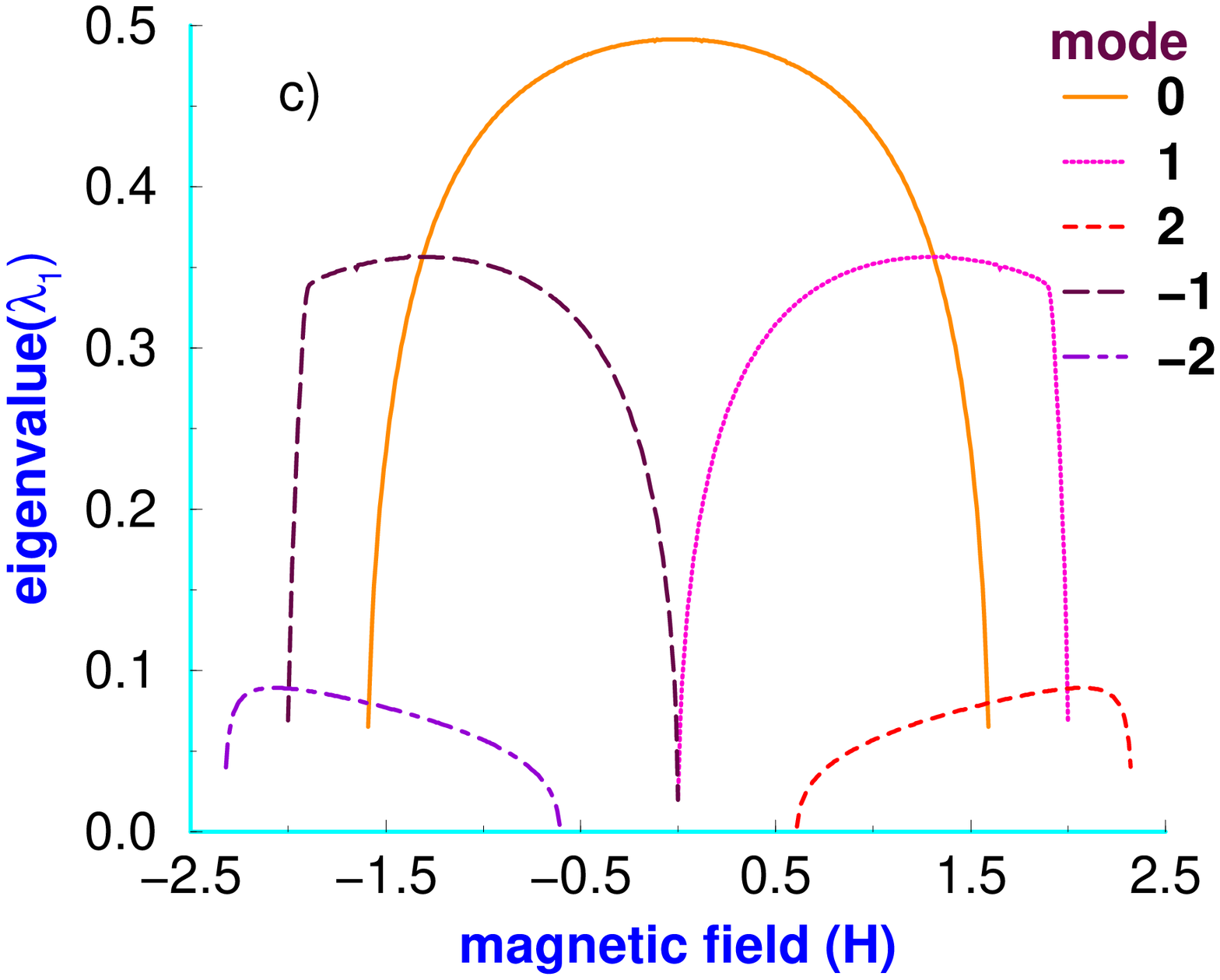,width=8.5cm,angle=0}}
 \caption{(a) Critical current $I_{max}$ and (b)
 magnetic flux at zero current, versus the magnetic field $H$, for the
 different modes.  $\ell=10$ and $D=1.4$.
 (c) The evolution of the
 lowest eigenvalue $\lambda_1$ with the external field for the
 different modes. At the extremes of each mode $\lambda_1$
 vanishes.}\label{fig2}\end{figure}

 \begin{figure}
 \centerline{\psfig{figure=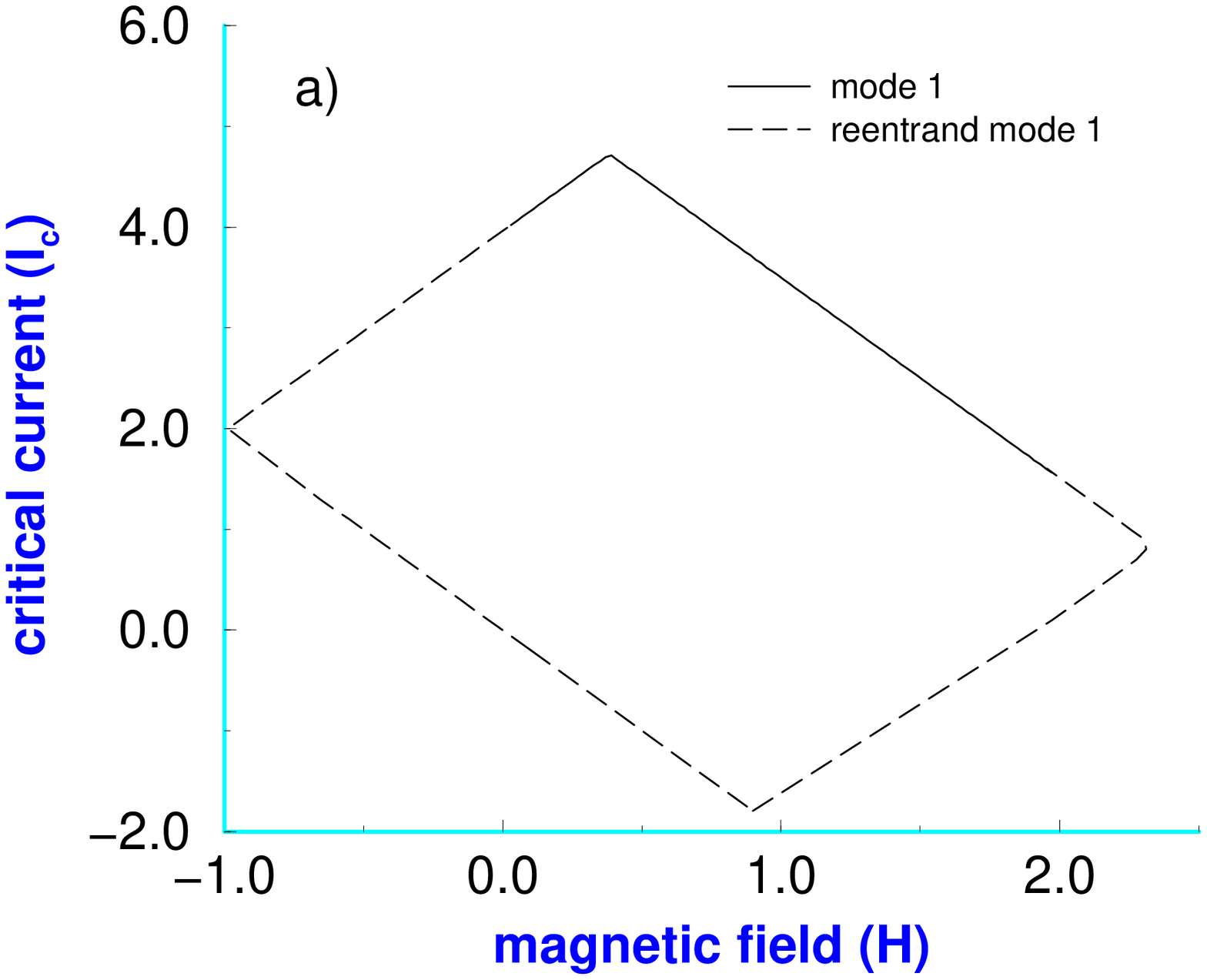,width=8.5cm,angle=0}}
 \centerline{\psfig{figure=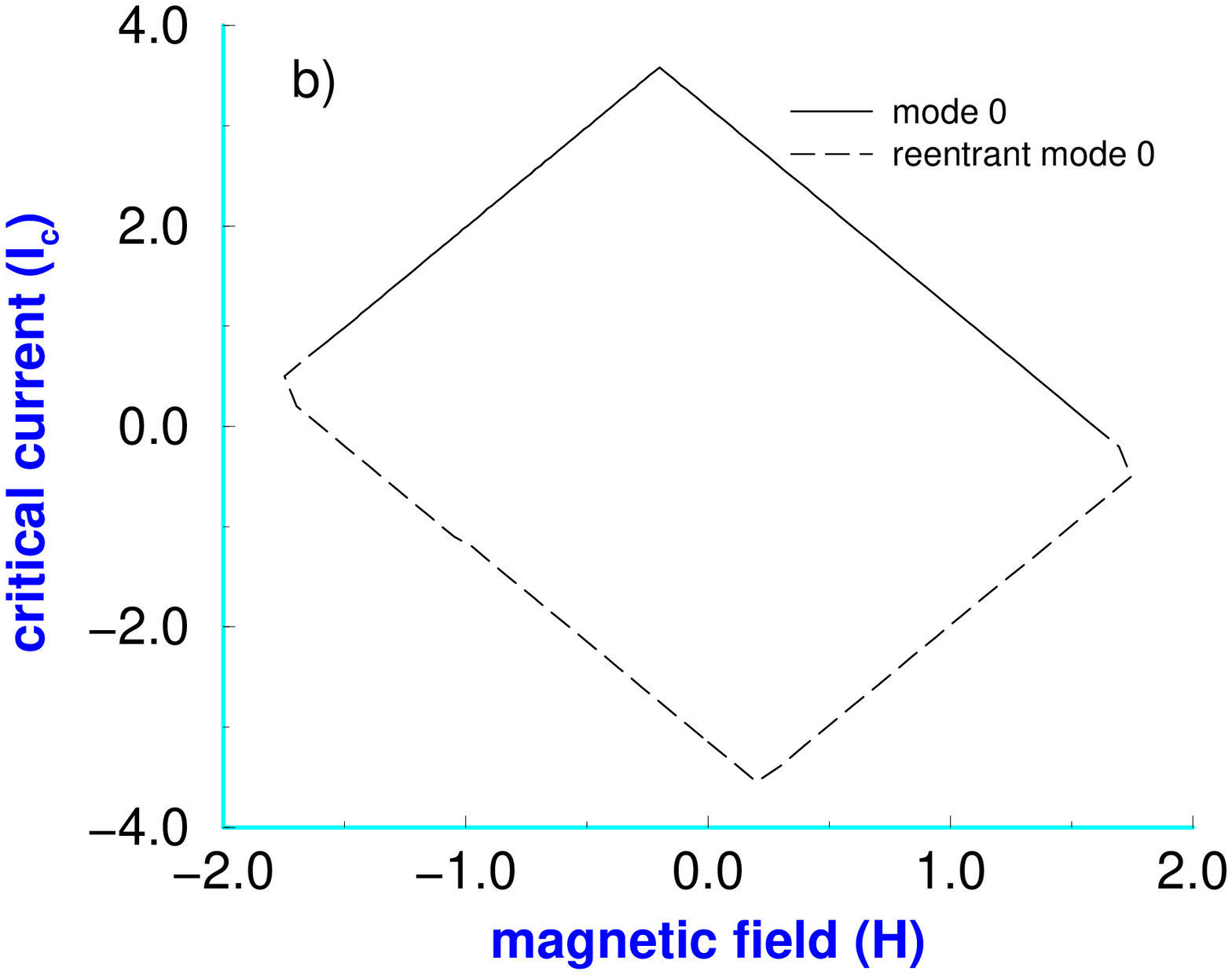,width=8.5cm,angle=0}}
 \caption{(a) Critical
 values of the bias current as a function of the magnetic field. The
 solid line is drawn for the mode 1 obtained with the usual procedure
 starting form zero current and increasing the current to the critical
 value, while the dashed line represents the values obtained with the
 reentrant procedure described in the text. (b) The same information as
 in (a), but for the mode 0. }\label{fig3}\end{figure}
 
 \begin{figure}
 \centerline{\psfig{figure=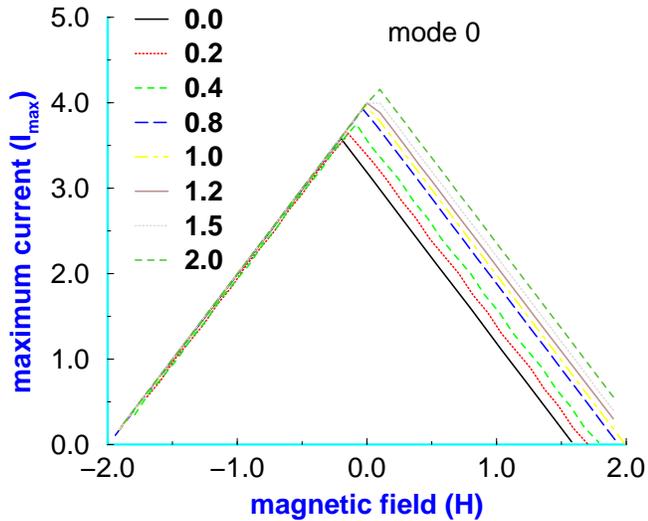,width=8.5cm,angle=0}} 
 \caption{Critical current as a function of the magnetic
 field, for the mode $0$, for different values of the defect critical
 current density $j_d$.}\label{fig4}\end{figure}

 \begin{figure}
 \centerline{\psfig{figure=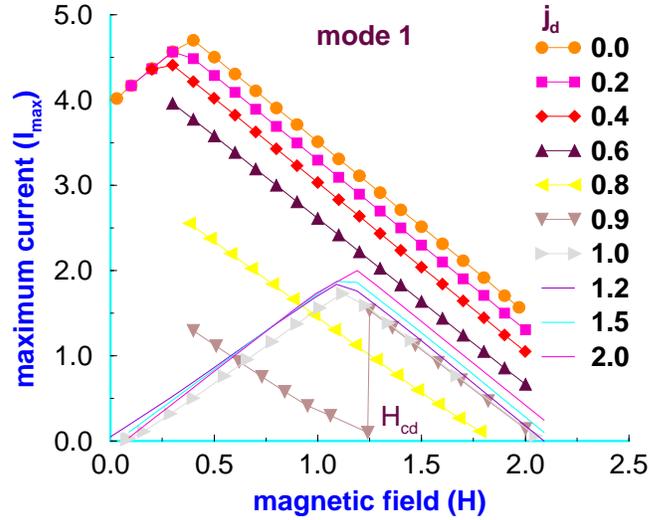,width=8.5cm,angle=0}} 
 \caption{Same as in Fig. 4, but for the mode $1$.}
 \label{fig5}\end{figure}

 \begin{figure}
 \centerline{\psfig{figure=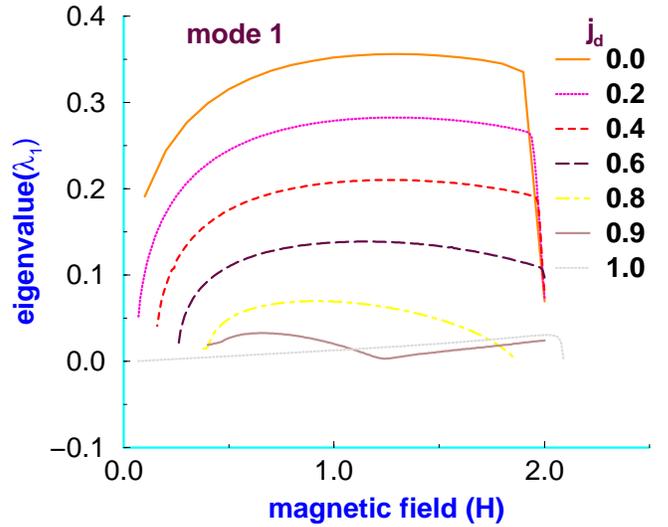,width=8.5cm,angle=0}}
 \caption{The lowest eigenvalue versus the magnetic
 field for the mode $1$ for different values of the defect critical
 current density $j_d$.}\label{fig6}\end{figure}

 \begin{figure}
 \centerline{\psfig{figure=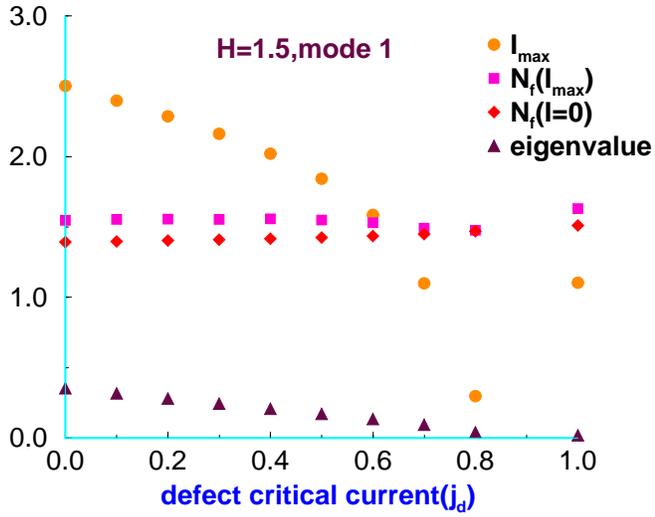,width=8.5cm,angle=0}}
 \caption{Critical current versus the defect critical
 current density $j_d$, for magnetic field equal to $H=1.5$, for the
 mode $1$. In the same graph the magnetic flux at zero and maximum
 current, and the lowest eigenvalue at $I=0$ are plotted as a function
 of the $j_d$.}\label{fig7} \end{figure}

 \begin{figure}
 \centerline{\psfig{figure=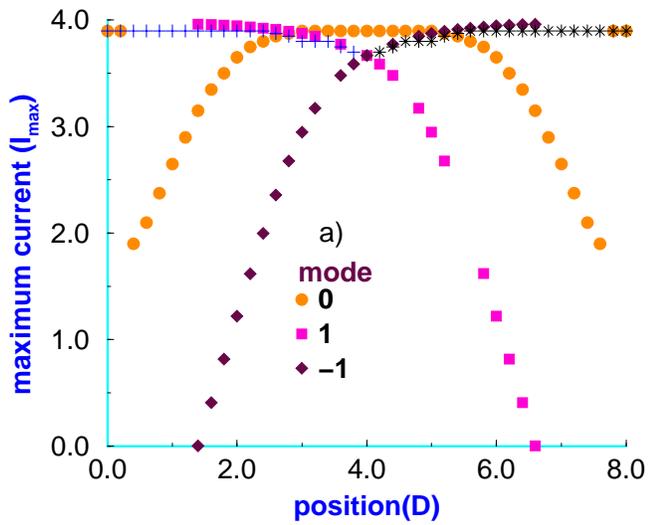,width=8.5cm,angle=0}}
 \centerline{\psfig{figure=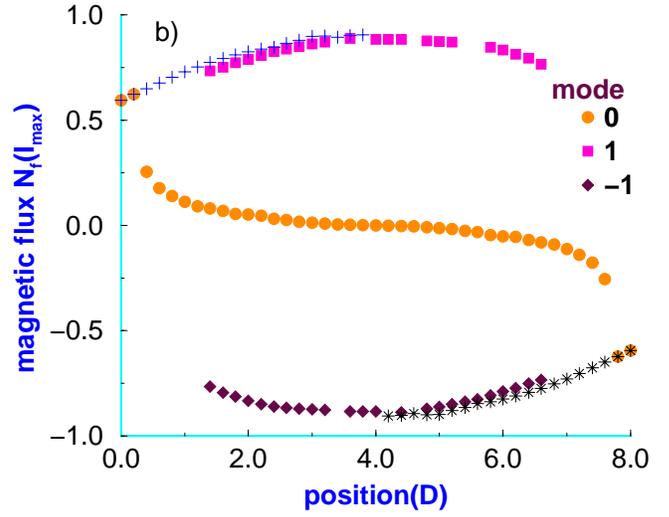,width=8.5cm,angle=0}}
 \centerline{\psfig{figure=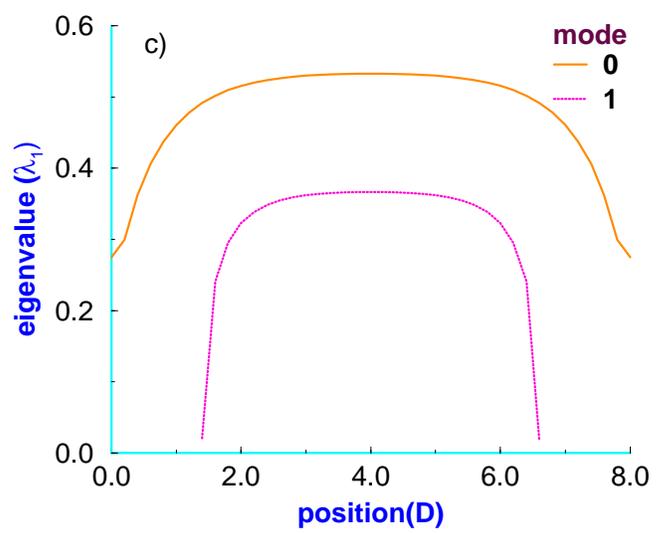,width=8.5cm,angle=0}}
 \caption{The  variation of (a) the maximum
 current $I_{max}$   and (b) the magnetic
 flux $N_f$ at the maximum current, versus the defect position $D$,
 measured from the right edge of the junction, for the modes $0$, $1$,
 $-1$. The crosses and stars lines are continuations of the two
 points at the two ends of the graph.  (c) the corresponding lowest
 eigenvalue at zero current versus the defect position $D$, for the
 modes $0$, $1$. The  $-1$ mode eigenvalue is the same as for the
 $1$ mode.}\label{fig8} \end{figure}

 \begin{figure}
 \centerline{\psfig{figure=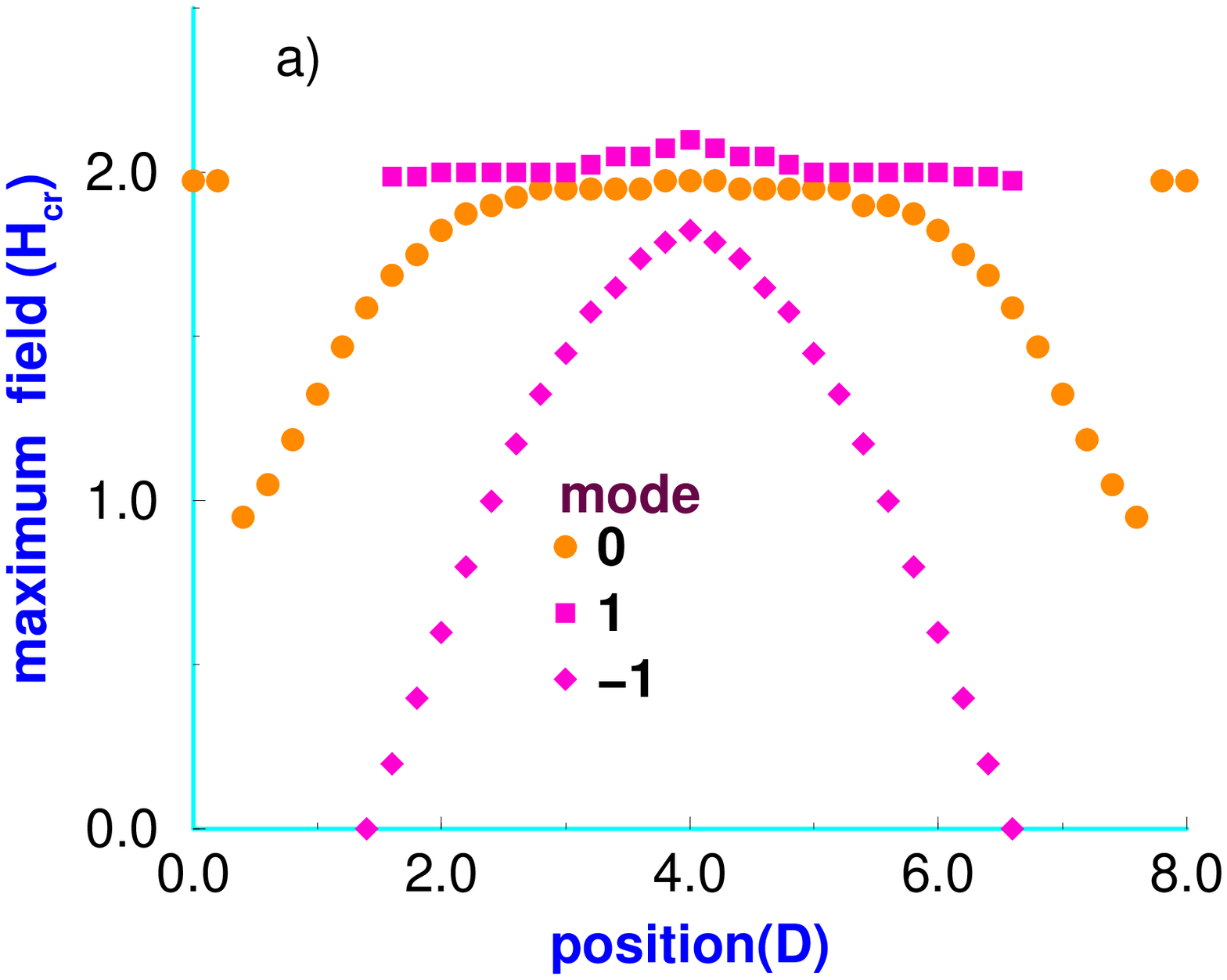,width=8.5cm,angle=0}}
 \centerline{\psfig{figure=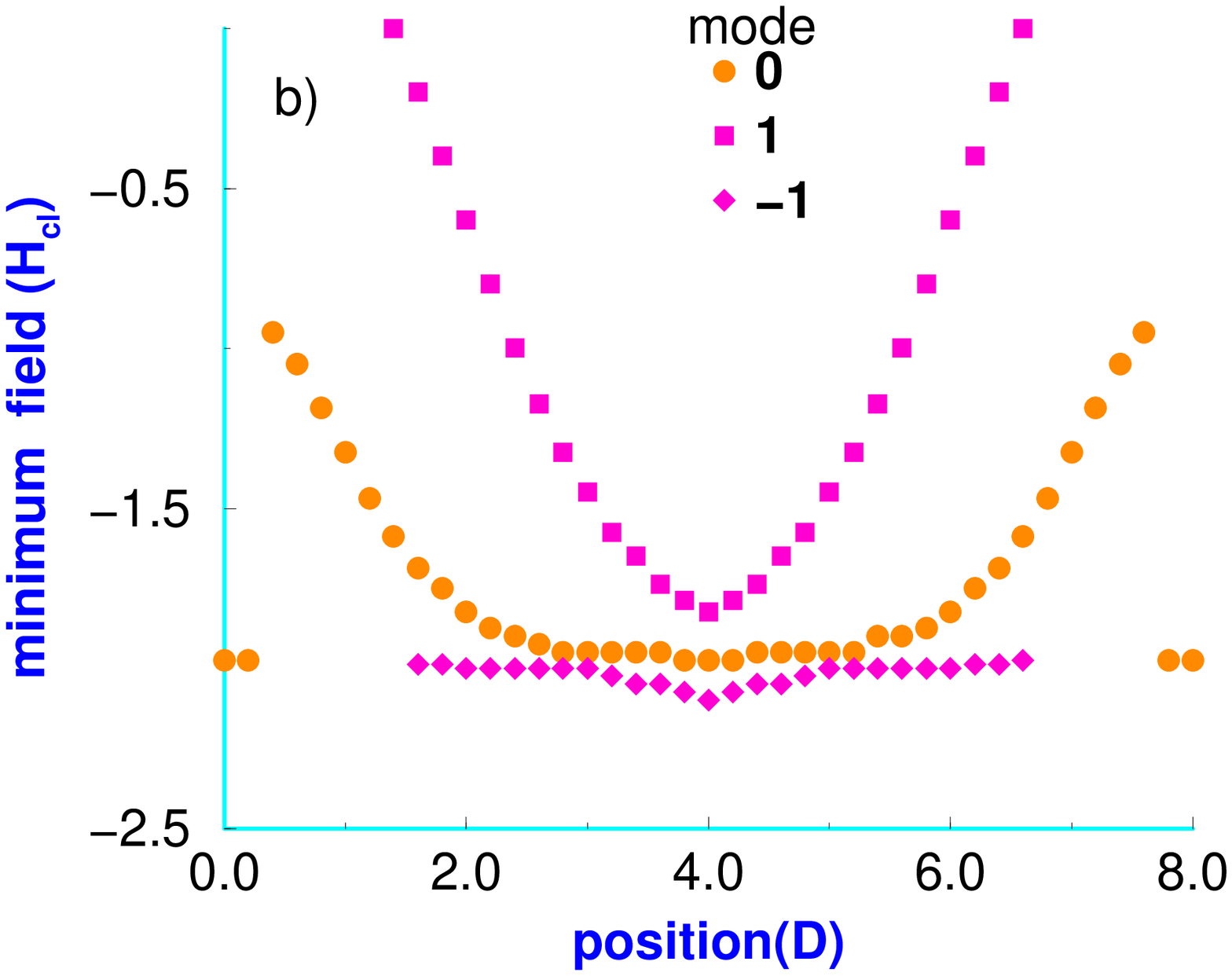,width=8.5cm,angle=0}}
 \caption{(a) The
 critical value of instability as we scan the magnetic field to the
 right$H_{cr}$ as a function of the defect position $D$, for the
 modes $0$, $1$, $-1$. (b) The same as (a) but to the left for the
 $H_{cl}$.}\label{fig9}\end{figure}
 
 \begin{figure}
 \centerline{\psfig{figure=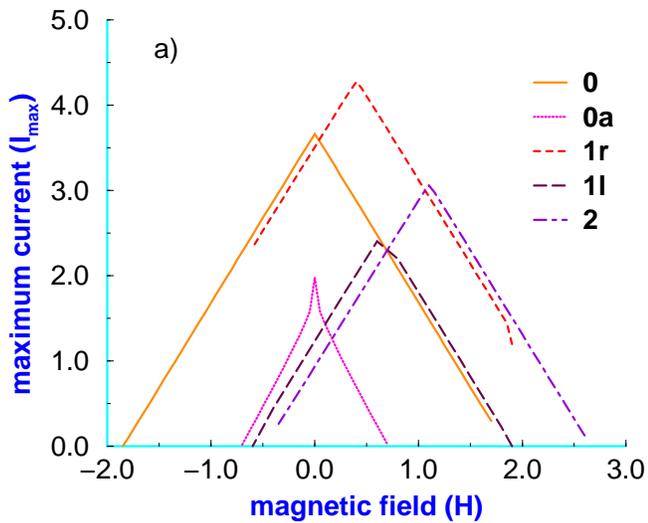,width=8.5cm,angle=0}}
 \centerline{\psfig{figure=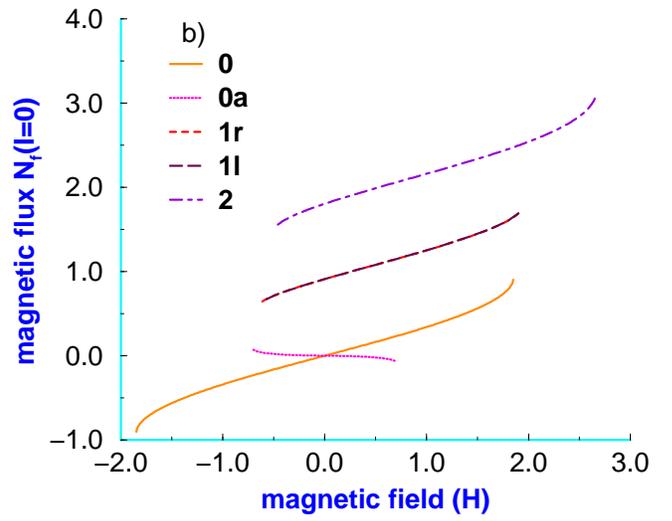,width=8.5cm,angle=0}}
 \caption{(a) Critical
 current $I_{max}$ and (b) magnetic flux $N_f$, versus the magnetic
 field $H$, for the different modes, for a junction of length
 $\ell=10$, which contains two symmetric pinning centers of length
 $d=2$.}\label{fig10} \end{figure}

 \begin{figure}
 \centerline{\psfig{figure=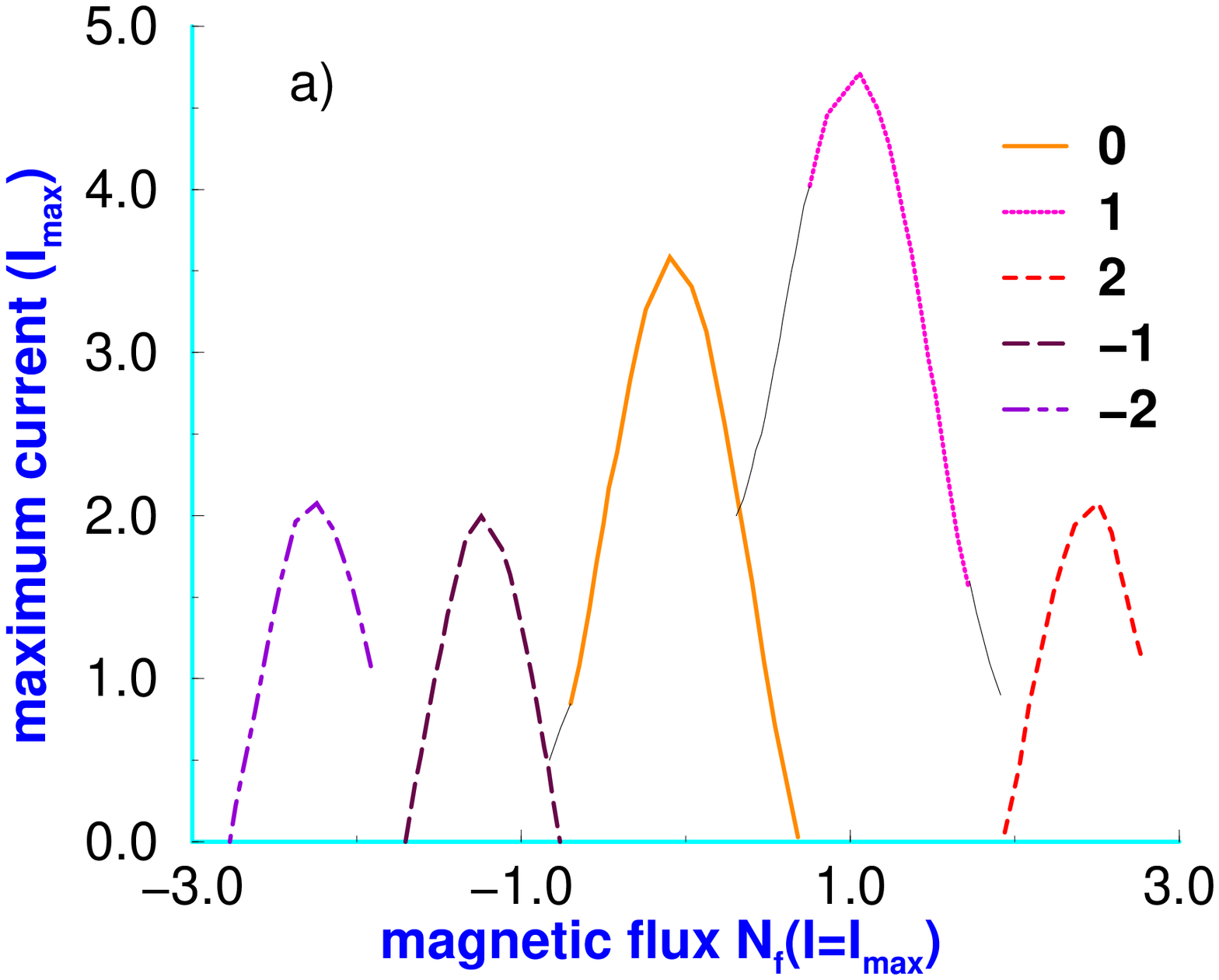,width=8.5cm,angle=0}}
 \centerline{\psfig{figure=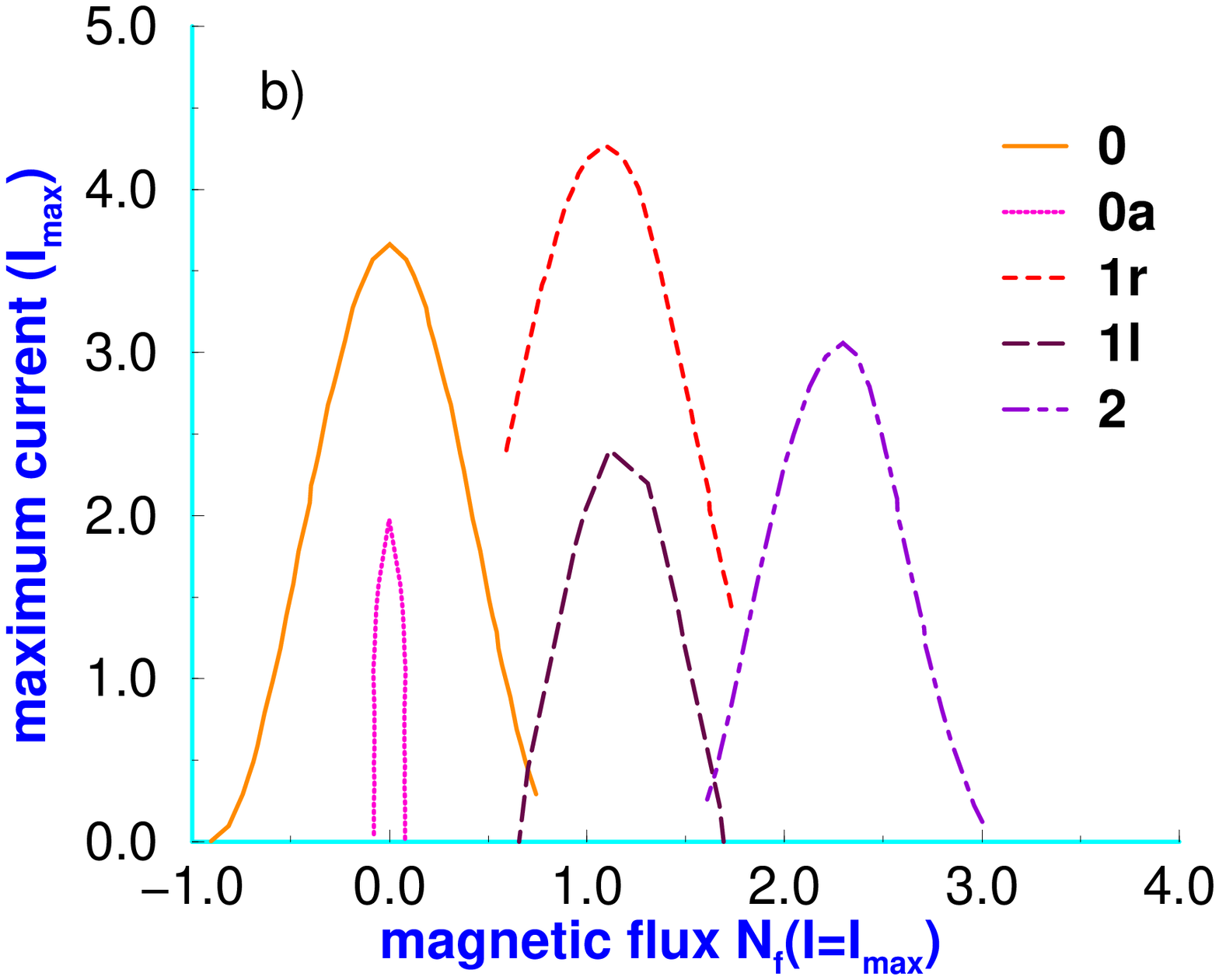,width=8.5cm,angle=0}}
 \caption{Critical
 current $I_c$, versus the magnetic flux $N_f$, at the maximum current
 for the different modes, for a junction of length $\ell=10$, (a) for
 the asymmetric defect case, and (b) for the two symmetric pinning
 centers of length $d=2$.}\label{fig11}\end{figure}
 
 \begin{figure}
 \centerline{\psfig{figure=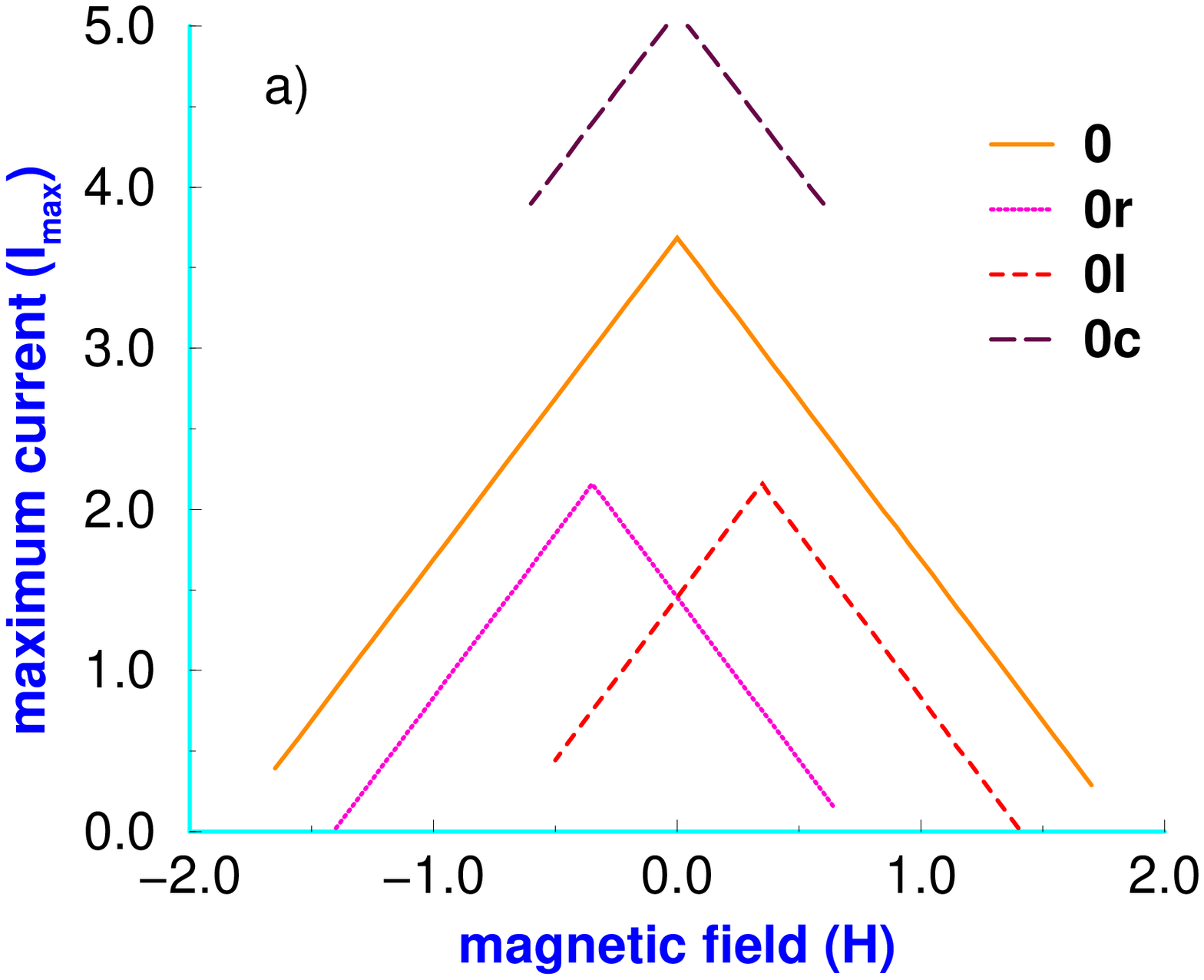,width=8.5cm,angle=0}}
 \centerline{\psfig{figure=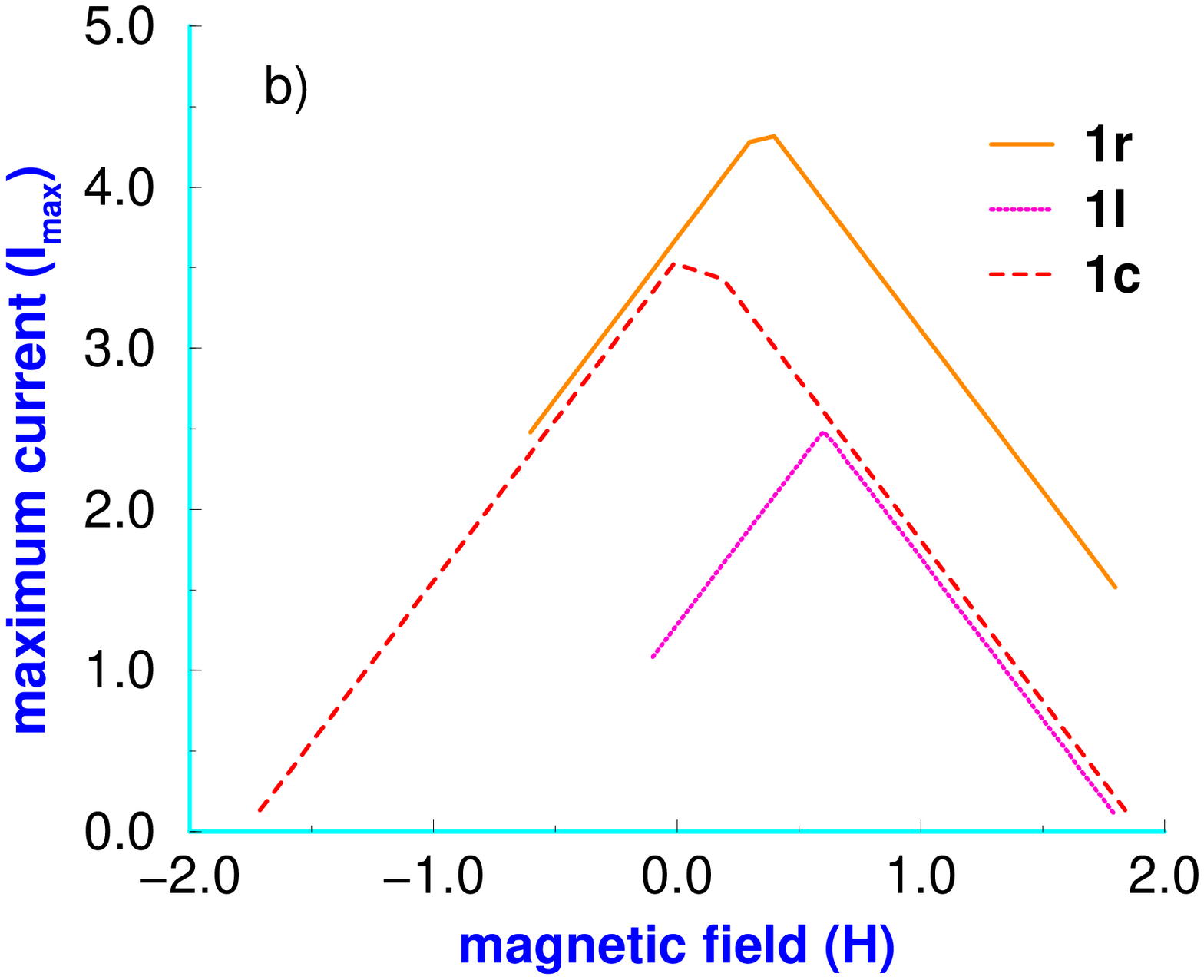,width=8.5cm,angle=0}}
 \centerline{\psfig{figure=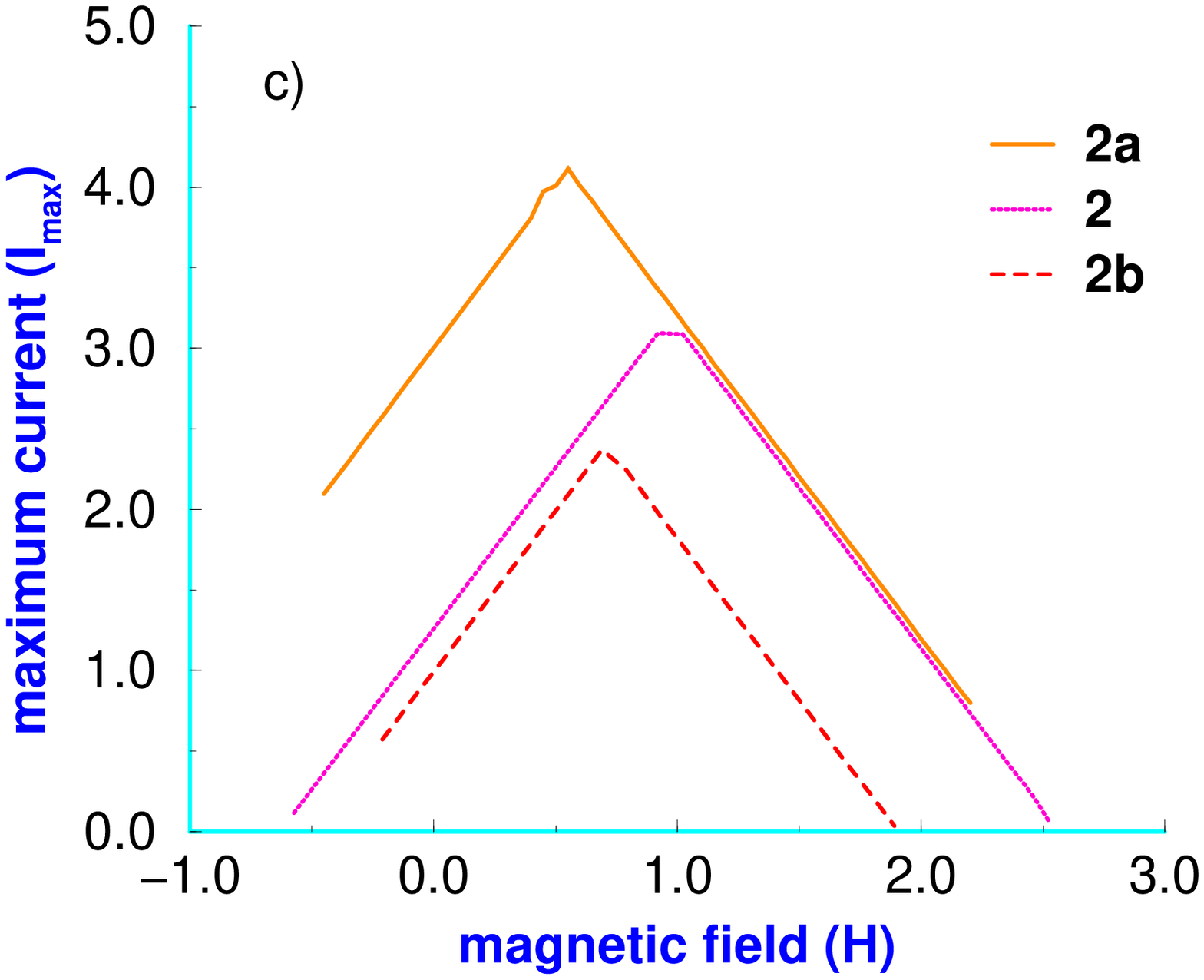,width=8.5cm,angle=0}}
 \centerline{\psfig{figure=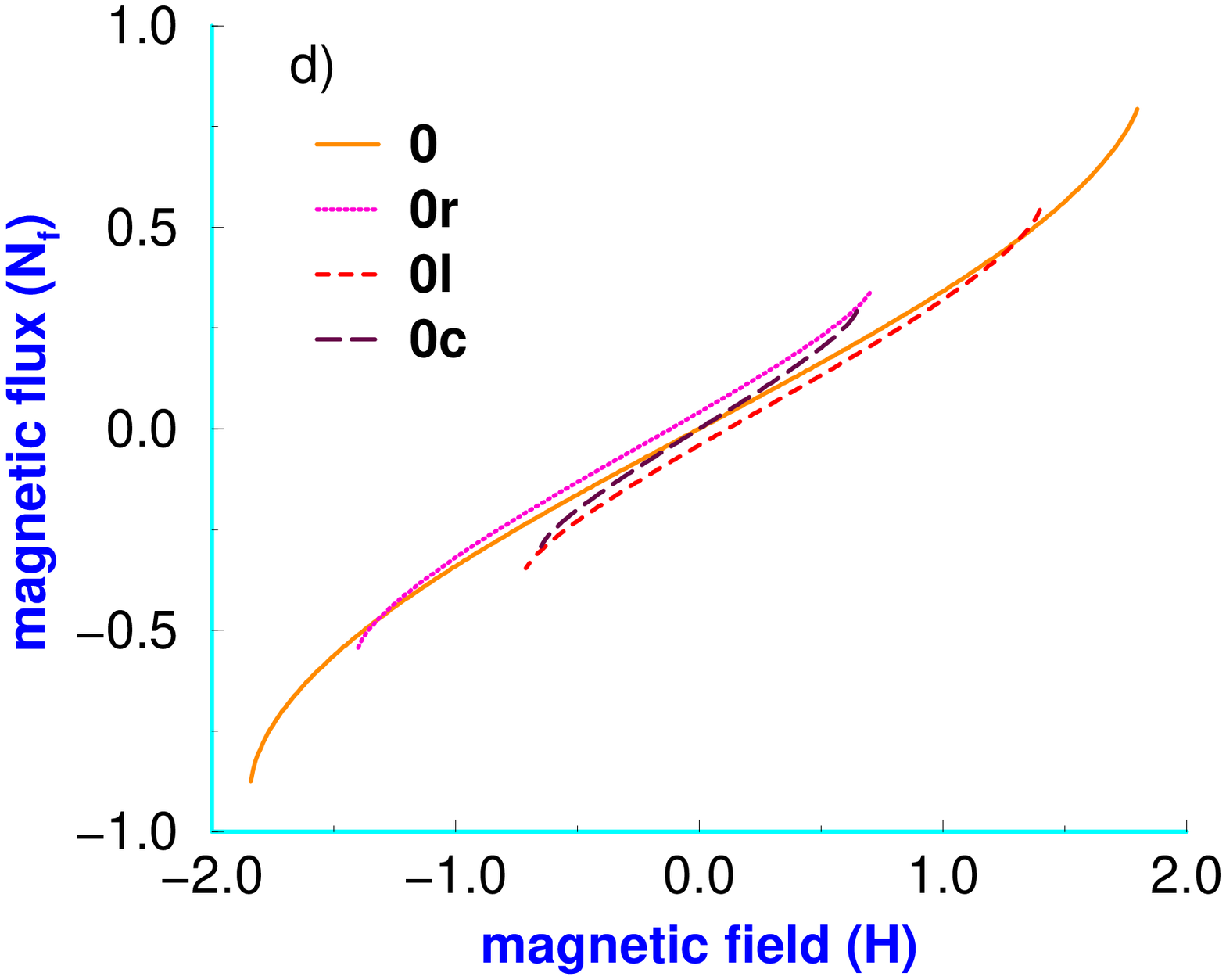,width=8.5cm,angle=0}}
 \centerline{\psfig{figure=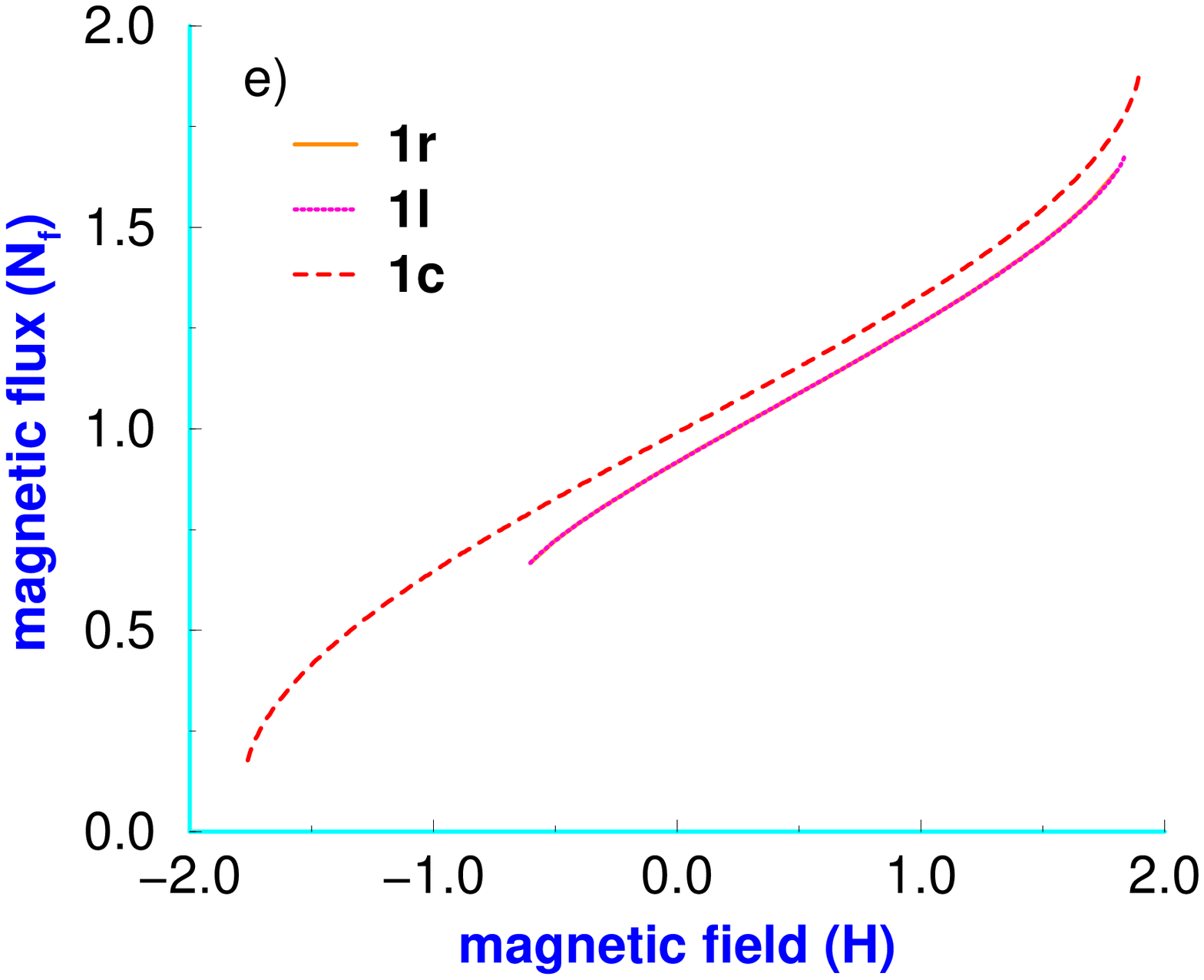,width=8.5cm,angle=0}}
 \centerline{\psfig{figure=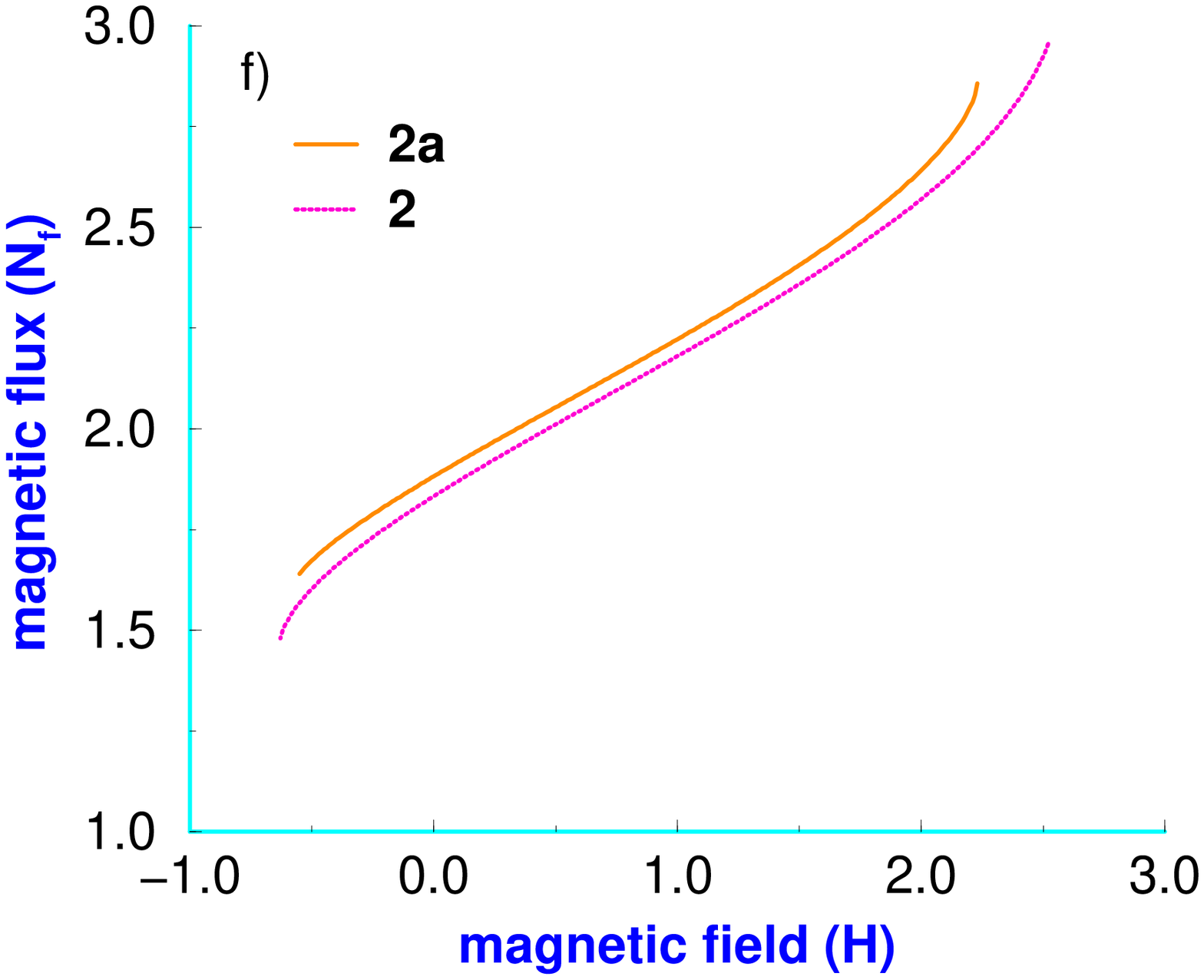,width=8.5cm,angle=0}}
 \caption{Critical current $I_{max}$ versus the magnetic
 field $H$, for the different modes (a) $0$, $0l$, $0r$, $0c$, (b)
 $1r$, $1l$, $1c$, and (c) $2$, $2a$, $2b$. for a junction of length
 $\ell=14.2$, which contains three symmetric pinning centers of length
 $d=2$. The corresponding magnetic flux is presented in  (d), (e),
 (f).}\label{fig12}\end{figure}
 
 \begin{figure}
 \centerline{\psfig{figure=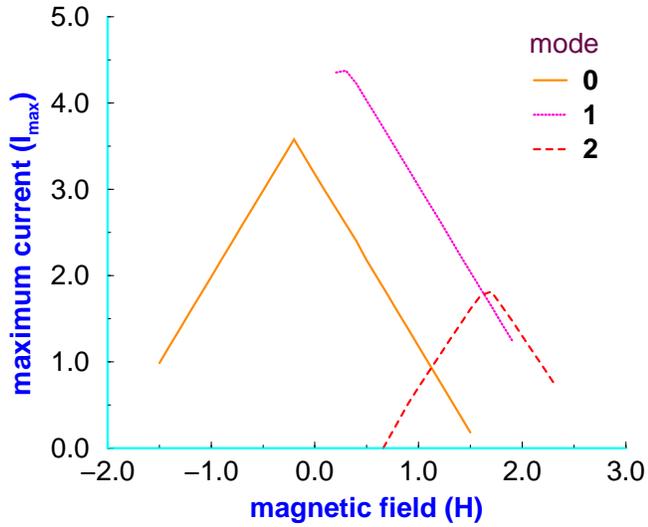,width=8.5cm,angle=0}} 
 \caption{Inline critical
 current $I_{max}$ versus the magnetic field $H$, for the
 modes $0$, $1$ and $2$,  for the junction with an
 asymmetric defect but smooth variation of the critical current
 density.$x_0=7.6$ and $\mu=2$.}\label{fig13}\end{figure}
 
 \begin{figure}
 \centerline{\psfig{figure=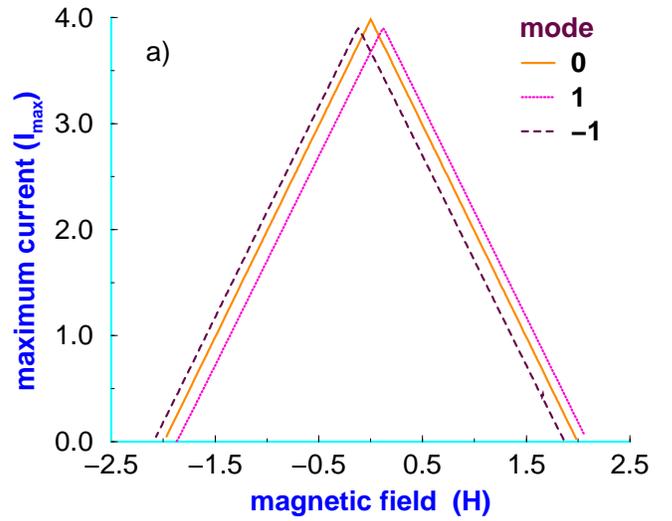,width=8.5cm,angle=0}}
 \centerline{\psfig{figure=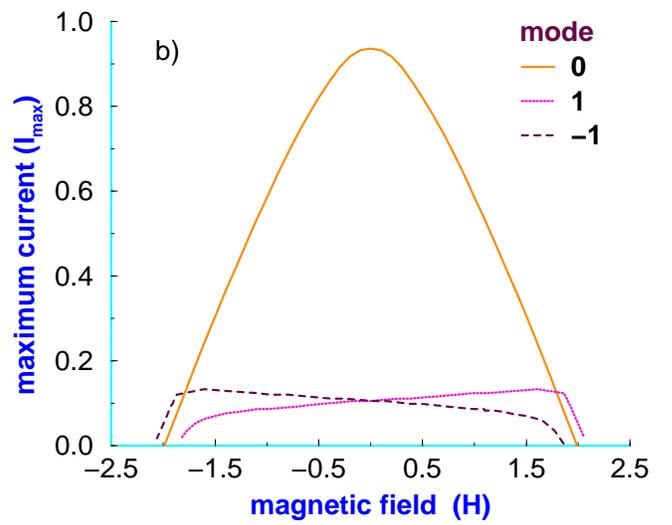,width=8.5cm,angle=0}}
 \caption{Critical current $I_{max}$ versus the magnetic
 field $H$, for the different modes, for (a) inline current and (b)
 overlap current, for the junction with a centered defect, and smooth
 variation of the critical current density.}\label{fig14}\end{figure}
 
 \begin{figure}
 \centerline{\psfig{figure=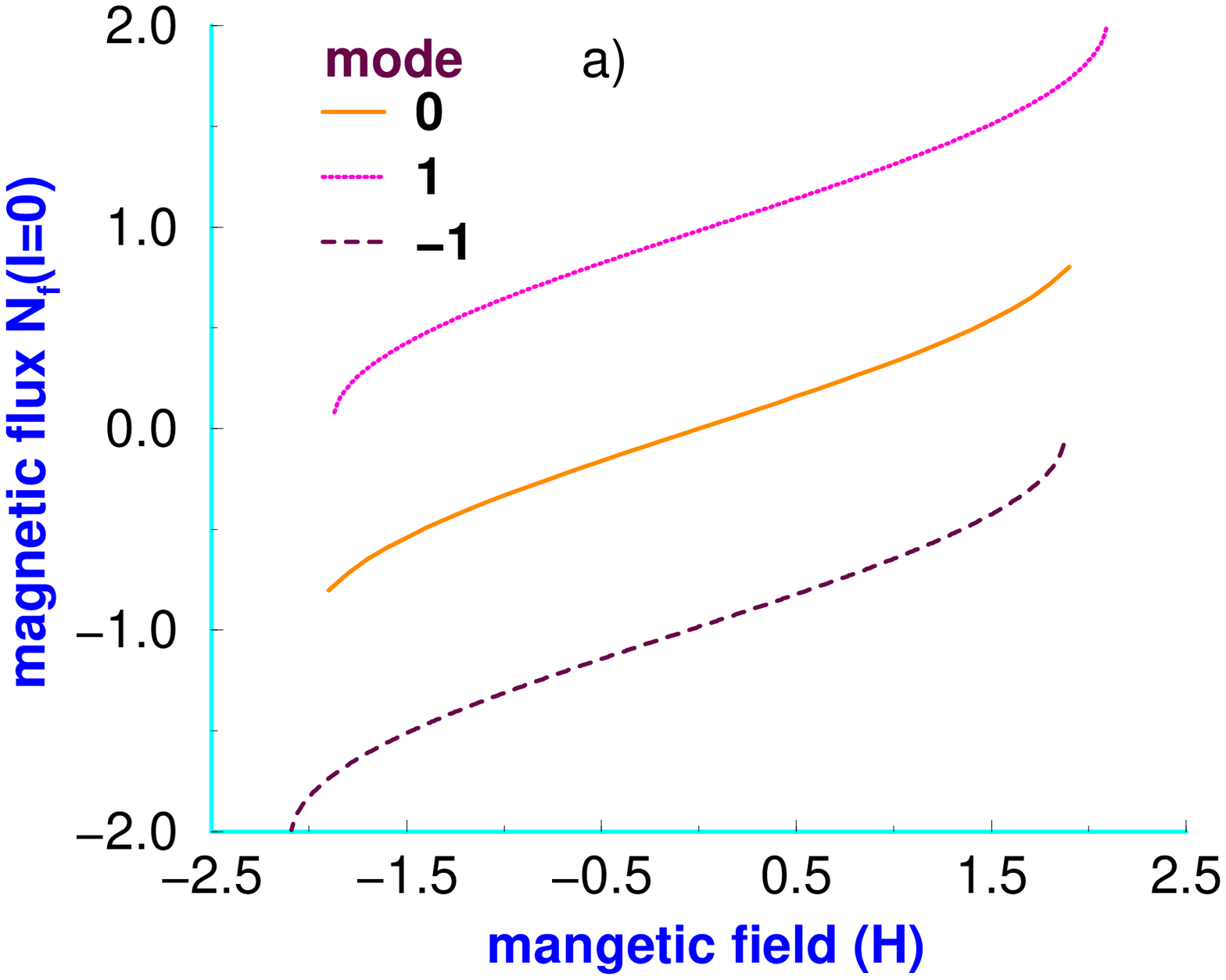,width=8.5cm,angle=0}}
 \centerline{\psfig{figure=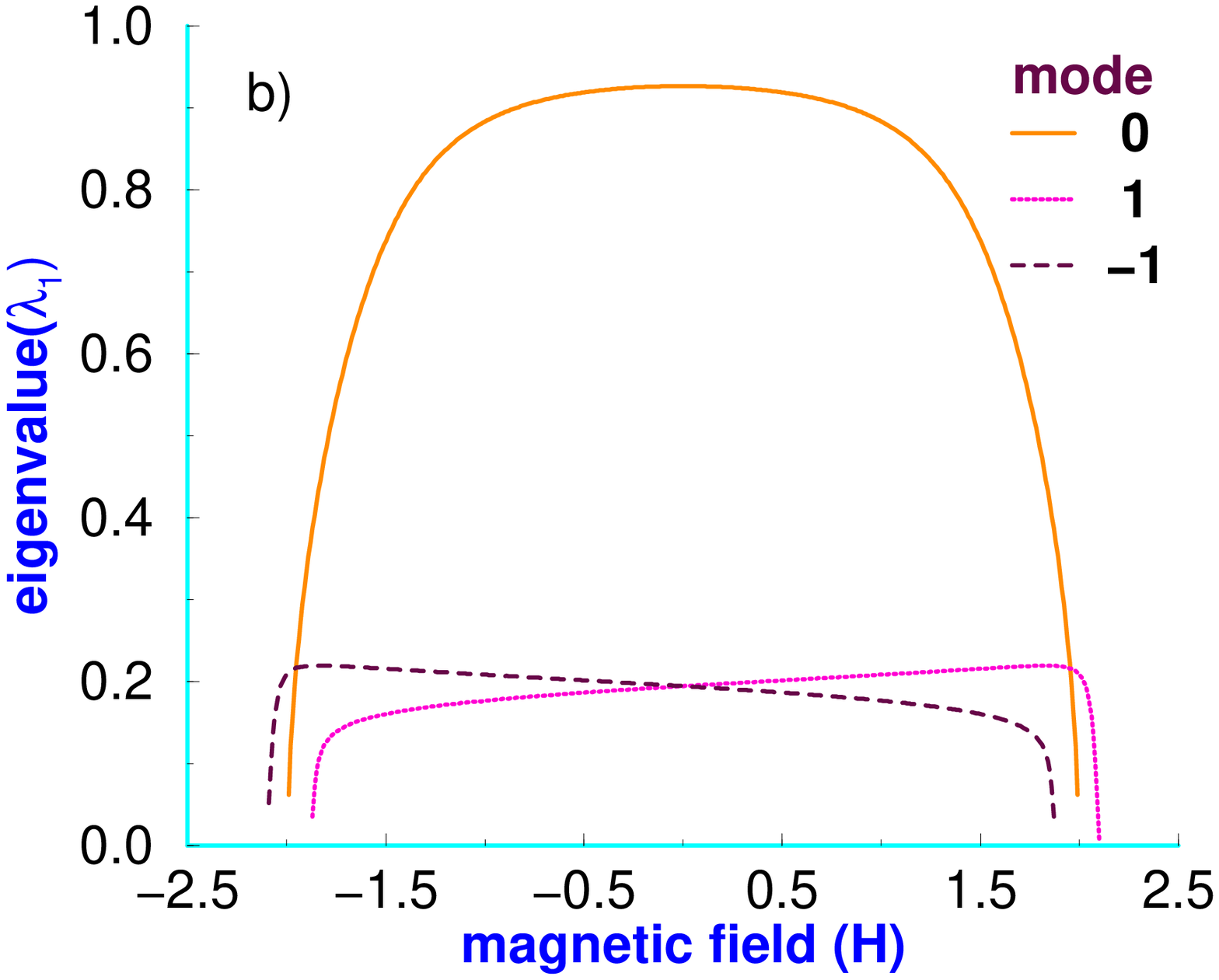,width=8.5cm,angle=0}}
 \caption{(a) Magnetic  flux at zero current
 as a function of the external field, for the  same
 type of inhomogeneity as in Fig.  14. (b)
 The  corresponding evolution of the lowest eigenvalue
 $\lambda_1$  for the different modes. At the
 end of each mode the $\lambda_1$ vanishes.}\label{fig15}\end{figure}
 
 \begin{figure}
 \centerline{\psfig{figure=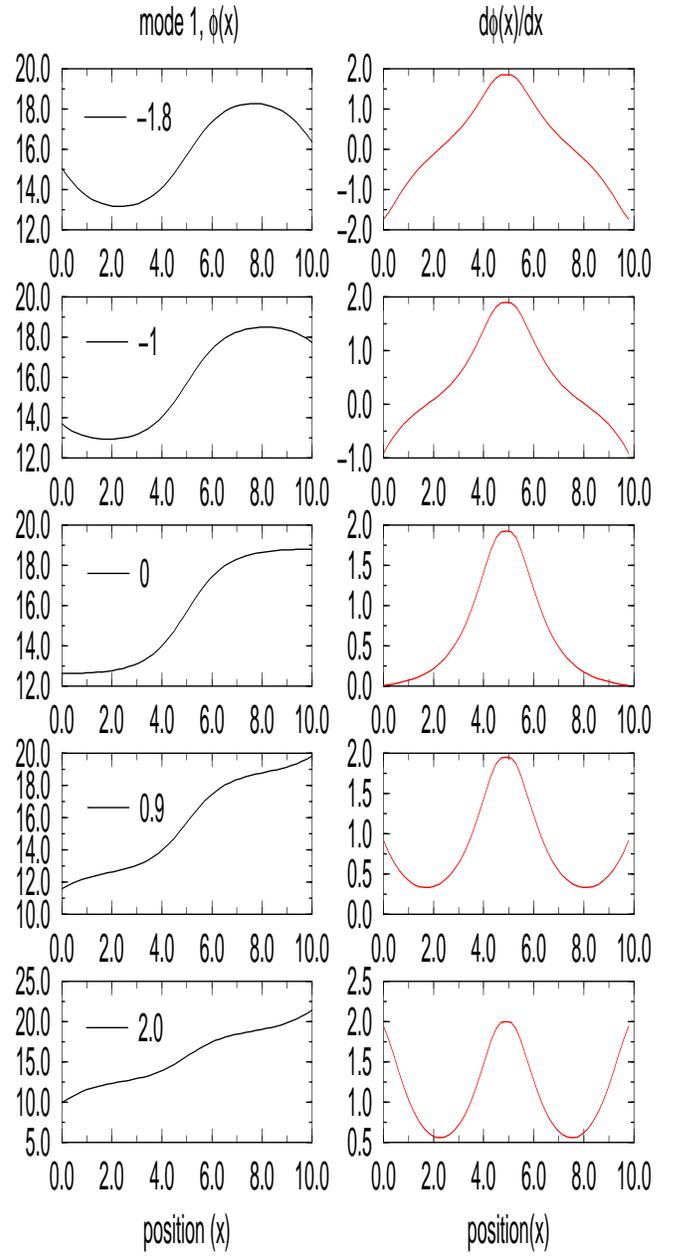,height=17cm,width=8.5cm,angle=0}}
 \caption{The  evolution of $\phi(x)$ and
 $\frac{d\phi(x)}{dx}$, for the mode $1$, as we  change the magnetic
 field at $I=0$, for the same type inhomogeneity  as in Fig. 14.
 Numbers are $H$ values. }\label{fig16}\end{figure} 
 
 \end{document}